\documentclass[twocolumn]{aastex631}

\usepackage{newtxtext,newtxmath,color}
\usepackage{amsmath, graphics, bm, graphicx}
\allowdisplaybreaks

\newcommand{\mathsym}[1]{{}}
\newcommand{\unicode}[1]{{}}

\newcommand{\ag}{\alpha}
\newcommand{\bg}{\beta}
\newcommand{\cg}{\gamma}

\newcommand{\dg}{\delta}
\newcommand{\Dg}{\Delta}
\newcommand{\kg}{\kappa}

\newcommand{\Om}{\Omega}
\newcommand{\om}{\omega}

\newcommand{\pd}{\partial}
\newcommand{\im}{{\rm i}}
\newcommand{\der}{{\rm d}}

\newcommand{\eps}{\varepsilon}

\newcommand{\bxi}{{\bm \xi}}

\newcommand{\e}{{\rm e}}

\newcommand{\he}{{\bm {\hat e}}}

\newcommand{\vphi}{\varphi}

\newcommand{\del}{\nabla}
\newcommand{\br}{{\bm r}}

\newcommand{\Ms}{M_\star}
\newcommand{\Rs}{R_\star}

\newcommand{\sg}{\sigma}

\newcommand{\ro}{{\rm orb}}

\newcommand{\Mjup}{{\rm M}_{\rm J}}

\newcommand{\Msun}{{\rm M}_\odot}

\newcommand{\Oms}{\Omega_\star}

\newcommand{\Mp}{M_{\rm p}}
\newcommand{\bJs}{{\bm J}_\star}
\newcommand{\hjs}{{\bm {\hat j}}_\star}
\newcommand{\bJo}{{\bm J}_{\rm orb}}
\newcommand{\hjo}{{\bm {\hat j}}_{\rm orb}}

\newcommand{\cO}{\mathcal{O}}

\newcommand{\btimes}{{\bm \times}}
\newcommand{\bcdot}{{\bm \cdot}}
\newcommand{\bdel}{{\bm \del}}

\newcommand{\be}{\begin{equation}}
\newcommand{\ee}{\end{equation}}

\begin{document}

\title{Damping Obliquities of Hot Jupiter Hosts by Resonance Locking}

\author[0000-0002-0786-7307]{J. J. Zanazzi}
\affiliation{
Astronomy Department, Theoretical Astrophysics Center, and Center for Integrative Planetary Science, University of California, Berkeley, \\
Berkeley, CA 94720, USA \\
}

\author[0000-0001-9420-5194]{Janosz Dewberry}
\affiliation{
Canadian Institute for Theoretical Astrophysics, 60 St. George Street, Toronto, ON M5S 3H8, Canada
}

\author[0000-0002-6246-2310]{Eugene Chiang}
\affiliation{
Astronomy Department, Theoretical Astrophysics Center, and Center for Integrative Planetary Science, University of California, Berkeley, \\
Berkeley, CA 94720, USA \\
}
\affiliation{
Department of Earth and Planetary Science, University of California, Berkeley, CA 94720, USA
}



\begin{abstract}
When orbiting hotter stars, hot Jupiters are often highly inclined relative to their host star equator planes. By contrast, hot Jupiters orbiting cooler stars are more aligned. Prior attempts to explain this correlation between stellar obliquity and effective temperature have proven problematic. We show how resonance locking --- the coupling of the planet’s orbit to a stellar gravity mode (g mode) --- can solve this mystery. Cooler stars with their radiative cores are more likely to be found with g-mode frequencies increased substantially by core hydrogen burning. Strong frequency evolution in resonance lock drives strong tidal evolution; locking to an axisymmetric g mode damps semi-major axes, eccentricities, and as we show for the first time, obliquities. Around cooler stars, hot Jupiters  evolve into spin-orbit alignment and may avoid engulfment. Hotter stars lack radiative cores, and therefore preserve congenital spin-orbit misalignments. We focus on resonance locks with axisymmetric modes, supplementing our technical results with simple physical interpretations, and show that non-axisymmetric modes also damp obliquity.  
Outstanding issues regarding the dissipation of tidally-excited modes and the disabling of resonance locks are discussed quantitatively.
\end{abstract}



\section{Introduction} 
\label{sec:intro}

\begin{figure*}
\centering
\includegraphics[width=\linewidth]{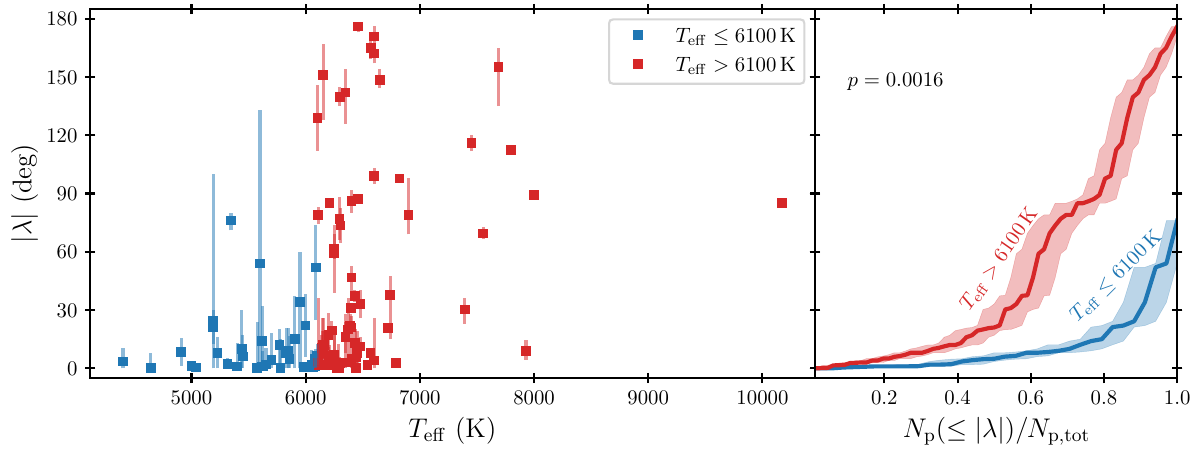}
\caption{
Projected stellar obliquities $\lambda$ (as distinct from deprojected or true 3D obliquities $\psi$) vs.~host star effective temperature $T_{\rm eff}$ of hot Jupiter systems (selecting for planet masses $M_{\rm p} \ge 0.3 \, {\rm M}_{\rm J}$ and semimajor axes $a \le 10\Rs$, where $\Rs$ is the host star radius; data from
\citealt{Albrecht+(2022), Rice+(2022), Rice+(2022b), Siegel+(2023), Espinoza-Retamal+(2023), Sedaghati+(2023), Hu+(2024)}).  
Obliquities are larger for hot high-mass stars (stellar effective temperature $T_{\rm eff} > 6100$ K, red points) than for cool low-mass stars (blue points).
This correlation \citep{Winn+(2010)} is statistically significant according to a two-sided Kolmogorov-Smirnov (KS) test, which yields a $p = 1.6 \times 10^{-3}$ probability that the cumulative obliquity distributions for cool and hot stars are drawn from the same underlying distribution (right panel). 
\label{fig:OblData}
}
\end{figure*}

\begin{figure}
\centering
\includegraphics[width=\linewidth]{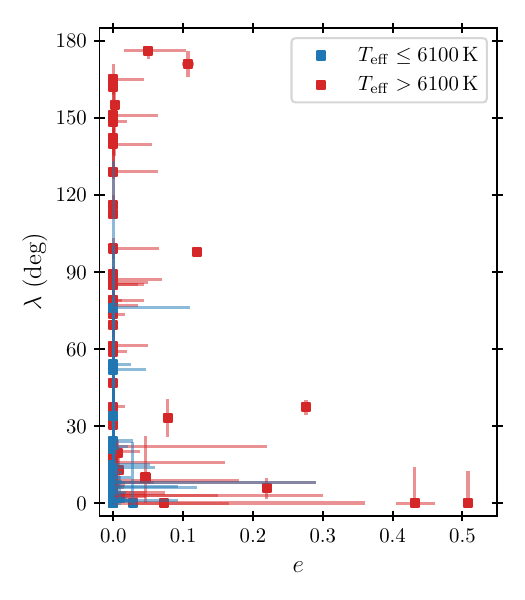}
\caption{
Projected obliquities vs.~eccentricities for hot Jupiters (identical sample to Fig.~\ref{fig:OblData}). 
Obliquities and eccentricities are lower for cool stars than for hot stars.
\label{fig:ecc_data}
}
\end{figure} 



\citet{Winn+(2010)} discovered that hot Jupiters orbiting cool stars have orbit normals that better align with stellar spin axes than hot Jupiters orbiting hot stars \citep[see also e.g.][]{Schlaufman(2010), Albrecht+(2012), Winn+(2017), MunozPerets(2018), Albrecht+(2021), HamerSchlaufman(2022), Rice+(2022), Rice+(2022b), Siegel+(2023)}. Figure \ref{fig:OblData} displays this correlation between stellar obliquity $\lambda$ and stellar effective temperature $T_{\rm eff}$. 
Figure \ref{fig:ecc_data} shows that $T_{\rm eff}$ seems to correlate also with orbital eccentricity $e$: cooler stars host more circular hot Jupiters.

High-eccentricity migration is one way to form hot Jupiters (e.g.~\citealt{DawsonJohnson(2018)}).  One imagines that giant planets are delivered from afar onto high-$e$, high-$\lambda$, low-periastron orbits  by, e.g., planet-planet scatterings or Lidov-Kozai oscillations. Presumably this initial delivery unfolds similarly around cool and hot stars. Subsequent tidal interactions with the star shrink the orbit by damping $e$ and orbital semi-major axis $a$. To reproduce the observations, tidal damping of $e$ and $\lambda$ would have to be, for some reason, more effective around cool stars than hot stars. In this paper we attempt to identify this reason. 


The dividing line used in Figs.~\ref{fig:OblData} and \ref{fig:ecc_data} to distinguish cool from hot stars is a stellar effective temperature $T_{\rm eff} = 6100$ K, a.k.a.~the Kraft break below which stars have convective envelopes and radiative cores, and above which stars have radiative envelopes and convective cores. Damping of the equilibrium tide, and by extension obliquity and eccentricity, has been argued to be more effective in turbulent convective envelopes, potentially explaining the $\lambda$-$T_{\rm eff}$ trend \citep{Winn+(2010), Albrecht+(2012)}. A problem with this idea is that convective eddy turnover times may be much too long compared to tidal forcing periods for turbulent viscosity to be significant \citep{GoldreichNicholson(1977), VidalBarker(2020)}. 
Even if convective dissipation (or the dissipation associated with the interaction between tidal flows and convection according to \citealt{Terquem(2021)}; but see \citealt{BarkerAstoul(2021)}) were somehow more effective, another problem with the equilibrium tide is that it results in wholesale semimajor axis decay. By the time the obliquity damps from equilibrium tidal dissipation, the star engulfs the planet \citep[see also][]{BarkerOgilvie(2009),Dawson(2014)}.

\cite{Lai(2012)} argued that dissipation of tidally excited inertial waves in convective zones could damp cool star obliquities while avoiding engulfment. 
One shortcoming of inertial wave dissipation is that it cannot take retrograde obliquities (which are observed for high-mass stars) and evolve them into prograde alignment 
\citep[e.g.][]{RogersLin(2013), ValsecchiRasio(2014), Xue+(2014), LiWinn(2016), Anderson+(2021), SpaldingWinn(2022)}. Another difficulty is that inertial wave dissipation scales with the host star's spin rate, which may be too slow on the main sequence to damp obliquities \citep[e.g.][]{LinOgilvie(2017),DamianiMathis(2018),SpaldingWinn(2022)}. Dissipation rates are greater on the pre-main sequence, but observations indicate high-$\lambda$ systems form late \citep{HamerSchlaufman(2022)}, consistent with  late-time dynamical instability in a high-$e$ migration scenario. Even if hot Jupiters formed early, high-mass stars have thick convective envelopes on the pre-main sequence, and inertial wave dissipation would predict their obliquities to be small, contrary to observation.

In this work we consider how $e$ and the (true 3D) obliquity $\psi$ evolve when the planet is resonantly locked with a stellar oscillation mode. The modes of interest are gravity modes (g modes), which exist in radiative (stably stratified) zones, not convective ones, and therefore behave differently between stars below and above the Kraft break. Gravity modes in the extensive radiative zones of stars have a dense frequency spectrum, and one can readily find a mode whose frequency matches (to within a low-integer factor) the planet's orbital frequency. In resonance lock \citep{WitteSavonije(1999), WitteSavonije(2001), Savonije(2008)}, the frequency match is preserved as a star evolves and its internal structure changes. Typically g-mode frequencies increase from increasing stratification due to hydrogen burning, and the planet's orbital frequency follows suit --- the planet migrates inward \citep{MaFuller(2021)}. Resonance locks and stellar evolution can change not only orbital semi-major axis $a$, but also eccentricity $e$ (e.g.~\citealt{Savonije(2008),Fuller(2017),ZanazziWu(2021)}). We show for the first time here that obliquity $\psi$ also evolves in resonance lock.

A full derivation of the equations governing $a$, $e$, and $\psi$ is given in appendix \ref{app:EL_tide_loss}; a condensed version sketching our physical picture and listing the main results is provided in section~\ref{sec:RLModel}. How g-mode frequencies evolve differently between cool and hot stars is explored with \texttt{MESA} stellar evolution calculations in section \ref{sec:GModeEv}. Results for the time evolution of $a$, $e$, and $\psi$ for proto-hot Jupiters in resonance lock are presented in section~\ref{sec:Results}. Section~\ref{sec:Disc} compares our theory with observations and offers some extensions, and section~\ref{sec:Conc} concludes. In appendix~\ref{app:Nonlin} we consider the still-uncertain physics of non-linear g-mode energy dissipation, and estimate the largest orbital distance out to which our theory might apply. 


\section{Resonance Locking Model}
\label{sec:RLModel}

\begin{figure}
\centering
\includegraphics[width=0.6\linewidth]{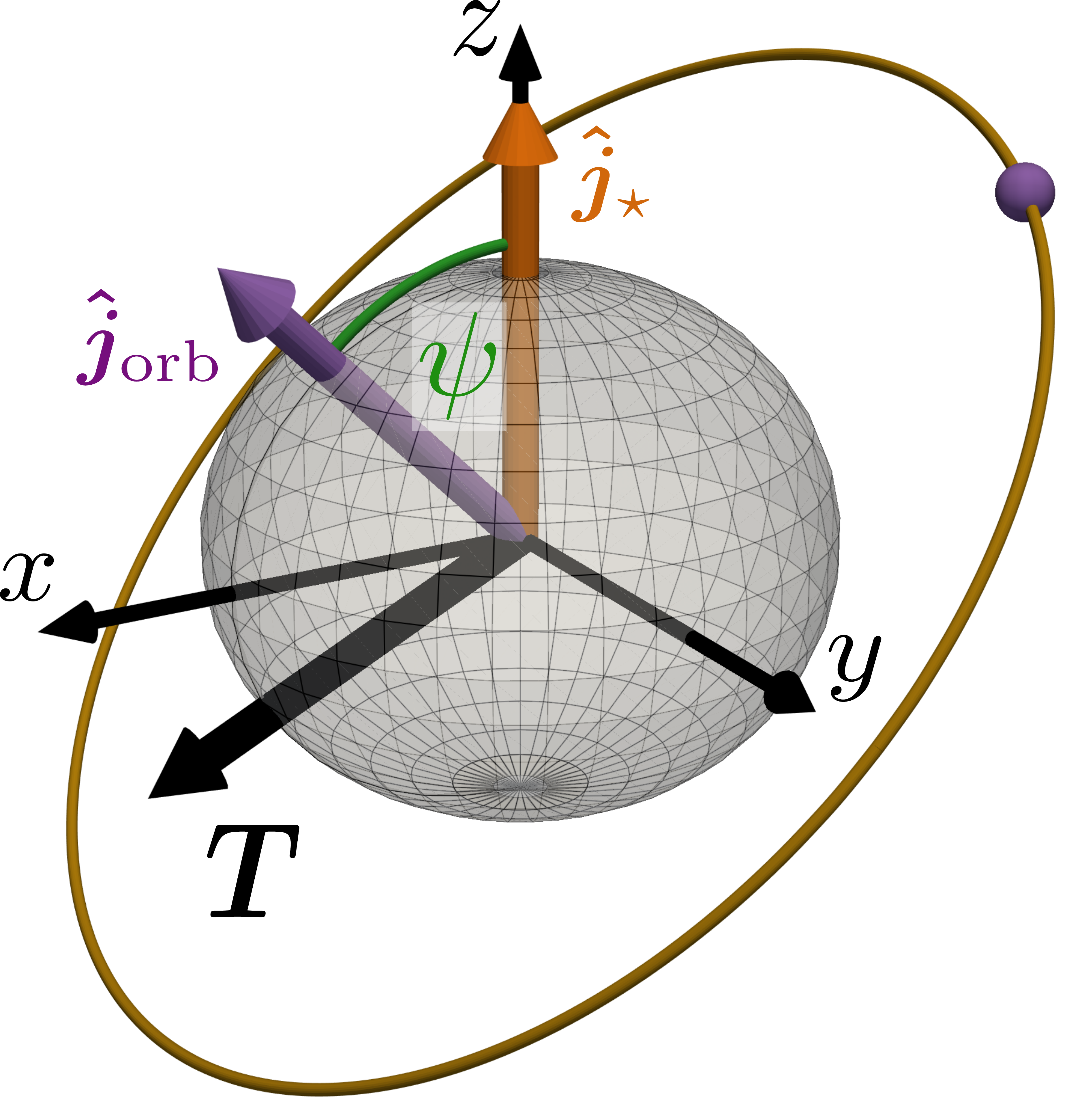}
\caption{
Coordinate system used to analyze the coupling between stellar spin and planet orbit. The $z$-axis is chosen to always point in the direction of the stellar spin $\hjs$, and the $y$-axis is chosen to always point normal to $\hjs$ and $\hjo$. The obliquity $\psi$ is the angle between $\hjs$ and $\hjo$. In this paper we consider only the components of the tidal torque ${\bm T}$ which act to change $\psi$, i.e.~only the $x$ and $z$-components.
\label{fig:setup}
}
\end{figure}

We begin with a general set of equations relating the angular momentum of the planet's orbit with the angular momentum of the stellar spin. The star has mass $\Ms$, radius $\Rs$, spin frequency $\Oms$, moment of inertia $I_\star = \kg_\star \Ms \Rs^2$, and spin angular momentum $\bJs = I_\star \Om_\star \hjs$ with $\hjs$ the unit vector parallel to the stellar spin axis. The star is orbited by a planet of mass $\Mp \ll \Ms$, having orbital semi-major axis $a$, eccentricity $e$, mean-motion $\Om = \sqrt{G \Ms/a^3}$, orbital energy $E_\ro = -G \Ms \Mp/(2a)$, and orbital angular momentum $\bJo = - (2 \sqrt{1-e^2} E_\ro/\Om)\hjo$, where $G$ is the gravitational constant and $\hjo$ is the unit orbit normal. The stellar obliquity $\psi$ is the angle between $\bJs$ and $\bJo$ ($\cos \psi = \hjo \bcdot \hjs$). We work in a Cartesian coordinate system such that $\hjs$ always lies in the $z$-direction, and $\hjo$ always lies in the $x$-$z$ plane (i.e.~$\hjs \btimes \hjo$ defines the $y$-direction). 
Energy is extracted from the orbit (from tidal dissipation) at a rate $\dot E_\ro < 0$. Angular momentum is exchanged (by tides) between planet and star at rate ${\bm T} = -\dot {\bm J}_\ro = +\dot {\bm J}_\star$; we restrict consideration to the torque's $x$ and $z$-components, 
${\bm T} = T_x \he_x + T_z \he_z$, since
$T_y$ just causes $\hjo$ to precess about $\hjs$ without changing $\psi$. We thus
have \citep{Lai(2012)}:
\begin{align}
    &\frac{1}{a} \frac{\der a}{\der t} = - \frac{\dot E_\ro}{E_\ro}, \label{eq:1} \\
        &\frac{\der J_\star}{\der t} = T_z, \label{eq:2} \\
    &\frac{\der J_\ro}{\der t} = -T_x \sin \psi - T_z \cos \psi, \label{eq:3} \\
    &\frac{\der \psi}{\der t} = - \frac{T_x}{J_\star} - \frac{T_x}{J_\ro} \cos \psi + \frac{T_z}{J_\ro} \sin \psi. \label{eq:4}
\end{align} 
See Figure~\ref{fig:setup} for an illustration of the coordinate system in which we are working.

\begin{figure*}
\centering
\includegraphics[width=0.8\linewidth]{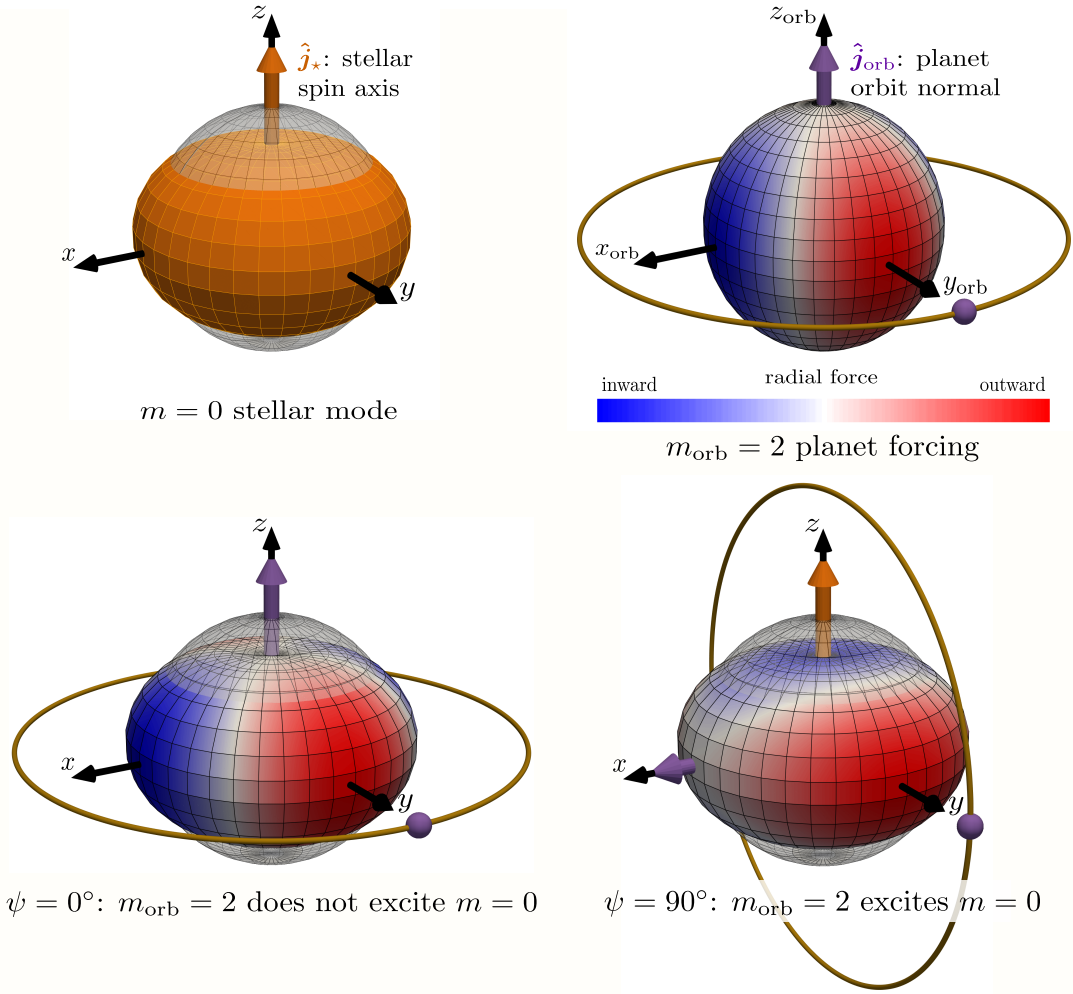}
\caption{An axisymmetric $m=0$ oscillation on the host star (top left) can be excited by the dominant $k=m_\ro=2$ component of the planet's tidal potential (top right) when   $\psi > 0$ (bottom right showing that the radial tidal force integrated over the stellar equator is net outward). When $\psi = 0$, tidal forces cancel over stellar azimuth and the $m=0$ mode cannot be excited (bottom left). 
\label{fig:ModeEx}
}
\end{figure*}

\begin{figure*}
\centering
\includegraphics[width=0.8\linewidth]{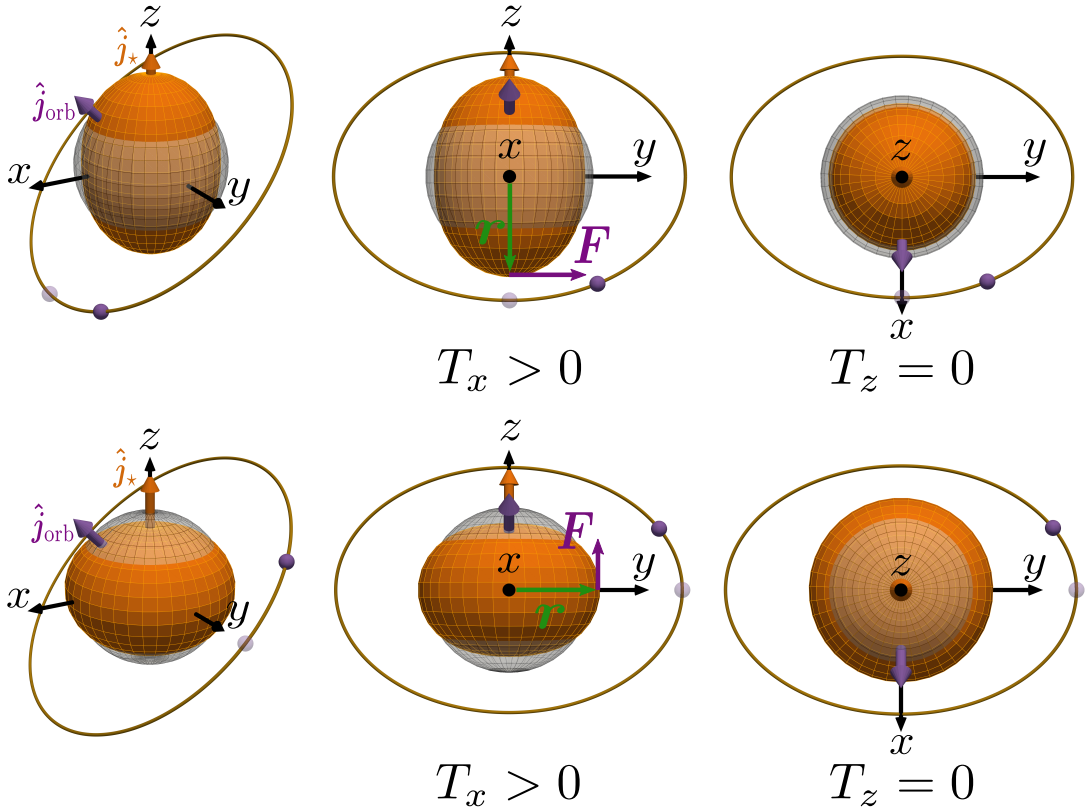}
\caption{
How a planet can tidally torque the host star into spin-orbit alignment, by way of an axisymmetric $m=0$ g-mode oscillation forced by the planet's $k = m_{\rm orb} = 2$ potential. The star stretches and compresses along its spin axis ($z$-axis) twice for every orbit of the planet. 
The solid purple sphere marks the planet's location which is displaced ahead of the star's tidal bulge due to tidal friction, while the see-through purple sphere displays the planet's location in the absence of tidal friction.
The top row shows different views of the same instant in time, when the star is maximally elongated but the planet is ahead at positive $y$. The bottom row shows a different time, when the star is maximally compressed and similarly lagged in phase relative to the planet. The tidal bulge at position vector ${\bm r}$ feels a force from the planet ${\bm F}$ and by extension a torque ${\bm T} = {\bm r} \times {\bm F}$ that is positive in the $x$-direction no matter the orbital phase (middle column). The torque $T_x > 0$ acts to bring the stellar spin direction ($\hjs$) toward the $x$-axis, closer to the orbital angular momentum direction ($\hjo$). Once spin and orbit are aligned, the torque vanishes (Fig.~\ref{fig:ModeEx}, bottom left panel).  
Because the oscillation is symmetric about the $z$-axis (right column), there is no torque in the $z$-direction.
\label{fig:torque}
}
\end{figure*}

Our theory is that $\dot E_\ro$ and ${\bm T}$ are driven predominantly by a particular oscillation mode in the star, resonantly excited by the planet. In resonance, an integer multiple $k$ of the planet's orbital frequency matches the stellar oscillation frequency:
\be
k \Om = \sg_\ag = \omega_\ag + m \Omega_\star
\ee
where $\sg_\ag$ is the mode oscillation frequency in the inertial frame, $\omega_\ag$ is the oscillation frequency in the frame rotating with the star, and $m$ is the mode's azimuthal number.
For most of this paper, we restrict attention for simplicity to ``zonal'' $m=0$ axisymmetric modes \citep{ZanazziWu(2021)}.

In resonance lock, oscillation and orbit track one another; the mode frequency evolves on the same timescale as the orbit evolves:
\begin{equation}
    t_{\rm ev}^{-1} \equiv \frac{\dot \sg_\ag}{\sg_\ag} = \eta_\ag \frac{\dot E_\ro}{E_\ro} 
    \label{eq:tev_def}
\end{equation}
where $\eta_\ag = 3/2$ for $m=0$ (see appendix \ref{app:EL_tide_loss} and \citealt{Fuller(2017)} for the case $m\neq 0$). In this paper, we assume the planet locks onto a stellar gravity mode (g mode), and compute how 
$\dot{\sg}_\ag/\sg_\ag$ evolves on the stellar main sequence (section \ref{sec:GModeEv}). 


In the inertial frame co-planar with the orbit, each term in the planet's tidal forcing potential varies as
\be
\exp \left( - \im k \Omega t 
+ \im m_\ro \vphi_\ro \right)
\ee
where $\vphi_\ro$ is the azimuthal coordinate measured in the orbital plane; note that $m_\ro$ differs from the stellar mode azimuthal number $m$ defined earlier. 
The dominant term in the forcing potential has $k=m_\ro=2$. Figure~\ref{fig:ModeEx} illustrates how $k=m_\ro=2$ tidal forcing can excite an $m=0$ mode in the star for $\psi \neq 0$ (bottom right panel), and conversely how $\psi = 0$ represents a fixed point where an $m=0$ mode is not excited (bottom left). Figure \ref{fig:torque} illustrates how this $m=0$ mode causes the obliquity to damp --- the planet exerts a torque on the tidally lagged stellar bulge that always acts to bring the stellar spin vector into closer alignment with the orbit normal, at all orbital phases of the planet. Only when spin and orbit align does the torque vanish. Thus, the $\psi=0$ fixed point is a stable attractor --- the obliquity under many initial conditions damps to zero.

We evaluate equations \eqref{eq:1}-\eqref{eq:4} for the case when planet and star are in resonance lock for $\{k,\ell,m\}=\{2,2,0\}$, where $\ell$ is the angular degree of the spherical harmonic representing the stellar oscillation mode (for context, most other studies of tidally forced oscillations focus on $\ell=2$, $m=2$). The full derivation is relegated to appendix \ref{app:EL_tide_loss}; the results are:
\begin{align}
    &\frac{1}{a} \frac{\der a}{\der t} = - \frac{1}{\eta_\ag t_{\rm ev}} \label{eq:7} \\
    &\frac{\der J_\star}{\der t} =  \ 
    0 \label{eq:8} \\ 
    &\frac{\der e}{\der t} = - \frac{1-e^2}{2e} \bigg(  1 - \frac{\sin \psi}{2\sqrt{1-e^2}}
    \tau  \bigg) \frac{1}{\eta_\ag t_{\rm ev}}
    \label{eq:9}\\
    &\frac{\der \psi}{\der t} = - \frac{1}{4 \sqrt{1-e^2}} \left( \frac{J_\ro}{J_\star} + \cos \psi \right) \tau \frac{1}{\eta_\ag t_{\rm ev}}. \label{eq:10}
\end{align}
Note that $J_\star$ does not change ($T_z = 0$; Fig.~\ref{fig:torque}, right column) because we are considering only axisymmetric stellar modes (see section \ref{subsec:nonaxi} and appendix \ref{app:EL_tide_loss} for a discussion of non-axisymmetric modes). The dimensionless torque $\tau = 2 \Omega T_x/\dot E$ is given by equation \eqref{eq:tau_def},
and depends on $e$ through Hansen coefficients 
$X^k_{\ell,m_\ro}$ 
and $\psi$ by a sum over 
Wigner-$d$ matrix coefficients (eq.~\ref{eq:Wigner_d}). When $\psi \ll 1$ and $e \gg \psi$, 
\begin{equation}
    \tau \simeq \frac{9}{8} \left( \frac{X^2_{2,2}}{X^2_{2,0}} \right)^2 \psi^3
\end{equation}
which vanishes as $\psi \to 0$. Equations~\eqref{eq:7} and~\eqref{eq:9} then imply $a(1-e^2) \simeq \text{constant}$, i.e.~the orbit circularizes along a constant angular momentum track  \citep{ZanazziWu(2021)}.  
In the opposite case when $e \ll 1$ and $\psi \gg e$,
 \begin{equation}
\tau \simeq \frac{2}{\sin \psi}
\label{eq:tau_smalle}
 \end{equation}
and the obliquity changes at a rate
\begin{equation}
    \frac{\der \psi}{\der t} \simeq -\left( \frac{J_\ro}{J_\star} + \cos \psi \right) \frac{1}{2\sin \psi} \frac{1}{\eta_\ag t_{\rm ev}}.
    \label{eq:obl_damp_rate}
\end{equation}
Obliquities damp if they are not too retrograde ($\cos \psi > -J_{\rm orb}/J_\star$). 
For hot Jupiters orbiting main-sequence stars with masses $\lesssim 1 \, \Msun$, 
obliquities typically damp 
because for such stars $J_{\rm orb}/J_\star > 1$ --- low-mass stars tend to rotate relatively slowly with 
$\sim$10-30 d
rotation periods, and have small $\kappa_\star \lesssim 0.15$ as they are centrally concentrated.




\newpage
\section{Gravity Mode Evolution Between Low-Mass and High-Mass Stars}
\label{sec:GModeEv}

\begin{figure}
\centering
\includegraphics[width=\linewidth]{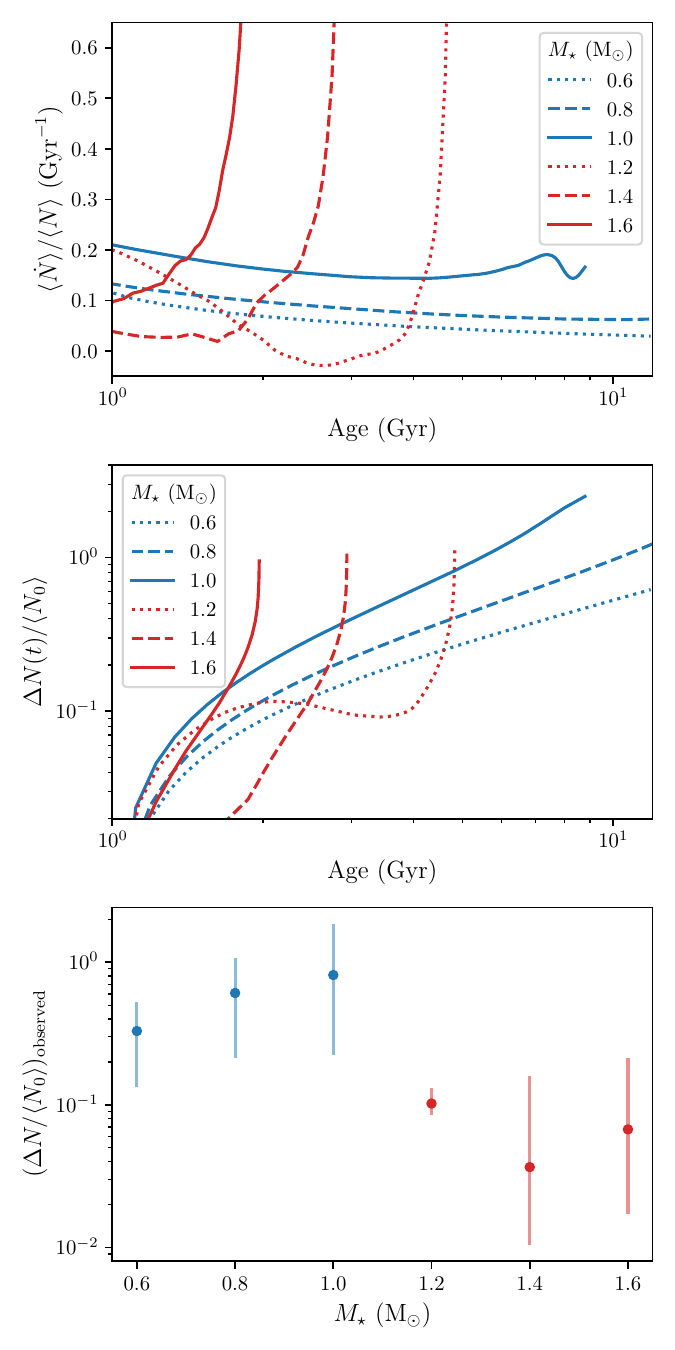}
\caption{
Rate of Brunt-V\"ais\"al\"a frequency evolution (top panel, eq.~\ref{eq:N_ave}) and accumulated frequency change (middle panel, eq.~\ref{eq:Porb_brunt}) vs.~main-sequence age, for stars of different masses as indicated. Although high-mass stars (red curves) increase their frequencies dramatically quickly toward the end of their hydrogen-burning lives (as their cores switch from being unstably to stably stratified), the speed-up phase is short-lived and consequently unlikely to be observed. Sampled over the entirety of their main-sequence lives, 
$\Delta N/ \langle N_0 \rangle$ is likely to be higher for low-mass stars, whose cores remain stably stratified throughout, than for high-mass stars (bottom panel showing median
 and $\pm 1 \sigma$ intervals, calculated by evaluating $\Dg N/N_0$ at 1000 uniformly spaced times over the interval $[t_0, t_{\rm max}]$).
\label{fig:aveN_t}
}
\end{figure}

In the low-frequency limit, g-mode frequencies in the host star's rotating frame are given by
\be
\om_\ag \simeq \frac{\sqrt{\ell(\ell+1)}}{\pi n_\ag} \langle N \rangle,
\label{eq:om_ag}
\ee
where $n_\ag \gg 1$ is the number of radial nodes, 
and the radial average of the Brunt-V\"ais\"al\"a frequency is given by
\be
\langle N \rangle = \int_{\rm rad} \frac{N}{r} \der r,
\label{eq:N_ave}
\ee
with the integral being taken over the stably stratified radiative zone ($N^2>0$).  As in \cite{ZanazziWu(2021)}, we focus on zonal modes (azimuthal number $m=0$), so $\om = \sg$, and $t_{\rm ev}$ is given by
\be
t_{\rm ev}^{-1} = \frac{\langle \dot N \rangle}{\langle N \rangle}
\label{eq:tev_brunt}
\ee
ignoring Coriolis forces.  

We define a fractional frequency change
\begin{equation}
\frac{\Delta N(t)}{\langle N_0 \rangle} \equiv \frac{\langle N (t)\rangle - \langle N_0 \rangle }{\langle N_0 \rangle} 
    \label{eq:Porb_brunt}
\end{equation}
and use the stellar evolution code \texttt{MESA} (release \texttt{r23.05.1}) to compute $\langle N(t) \rangle$ from $t_0 \equiv 1 \, {\rm Gyr}$ (the subscript 0 denotes evaluation at time $t_0$) to $t_{\rm max} = \min(t_{\rm MS}, 12 \, {\rm Gyr})$, with the 
end of the 
main-sequence lifetime $t_{\rm MS}$ defined by when the star's core 
has depleted its hydrogen. The \texttt{MESA} inlists we use are identical to those in \cite{Fuller(2017)}. How our results depend on $t_0$ and $t_{\rm max}$ is discussed in section \ref{subsec:t0}.

\begin{figure*}
\centering
\includegraphics[width=\linewidth]{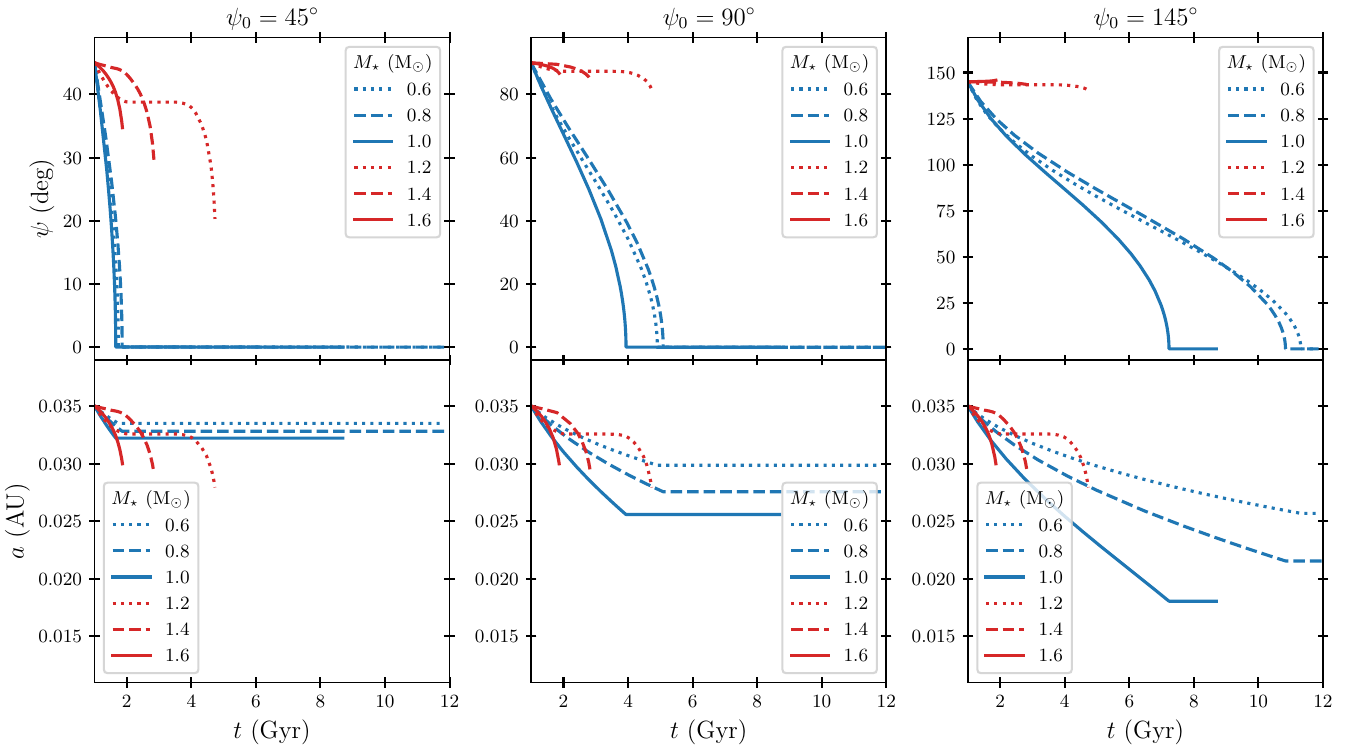}
\caption{
Orbital evolution of hot Jupiters in resonance lock with a stellar gravity mode ($\{k,l,m\} = \{2,2,0\}$), integrated from $t_0 = 1$ Gyr through the stellar main-sequence lifetime or $t = 12$ Gyr, whichever comes first. Here planet orbits are assumed circular for all time. Obliquities damp more readily for cool stars (blue curves) than hot stars (red curves). For hot stars with $\psi_0=145^\circ$, obliquities actually increase slightly, a consequence of their having $J_{\rm orb}/J_{\star} < 1$ (unlike for cool stars).
\label{fig:orb_ev_circ}
}
\end{figure*}

Over the star's main sequence lifetime, in response to the loss of gas pressure from the conversion of hydrogen into helium and the increase in mean molecular weight, the core contracts and becomes more dense and stratified. Thus $\langle N \rangle$ tends to increase, the more so when the region of stable stratification extends down to the star's core, as it typically does for cool but not for hot stars. Figure \ref{fig:aveN_t} illustrates how the evolution of the star-averaged Brunt-V\"ais\"al\"a frequency differs between cool low-mass stars (blue curves) and hot high-mass stars (red curves). The sign of the difference can be subtle and varies with context. The top panel of Fig.~\ref{fig:aveN_t} shows that for high-mass stars at the end of their main-sequence lifetimes, the mode evolution rate $\langle \dot{N} \rangle / \langle N \rangle$ actually exceeds that for low-mass stars --- see how the red curves ultimately skyrocket above the blue curves. The last-minute speed up of mode evolution in high-mass stars, which occurs as they ascend the sub-giant branch and their cores switch from being convective to radiative, is so great that even though low-mass stars live longer than high-mass stars, the total change $\Delta N$ accumulated over main-sequence lifetimes is comparable between low-mass and high-mass stars (middle panel of Fig.~\ref{fig:aveN_t}). Nevertheless, because the mode frequency speed-up for high-mass stars occurs over a small fraction of the high-mass main-sequence lifetime, the time-averaged value for $\Delta N$ is actually lower for high-mass as compared to low-mass stars (bottom panel of Fig.~\ref{fig:aveN_t}). The upshot is that given a low-mass main-sequence star and a high-mass main sequence star drawn randomly from the field, the low-mass star is more likely to have experienced a larger fractional change in its g-mode frequency. The expected change is on the order of unity for low-mass stars, and  $\lesssim 10\%$ for high-mass stars. This difference underpins our theory that stellar obliquities have evolved more for low-mass than high-mass stars.

\begin{figure*}
\centering
\includegraphics[width=\linewidth]{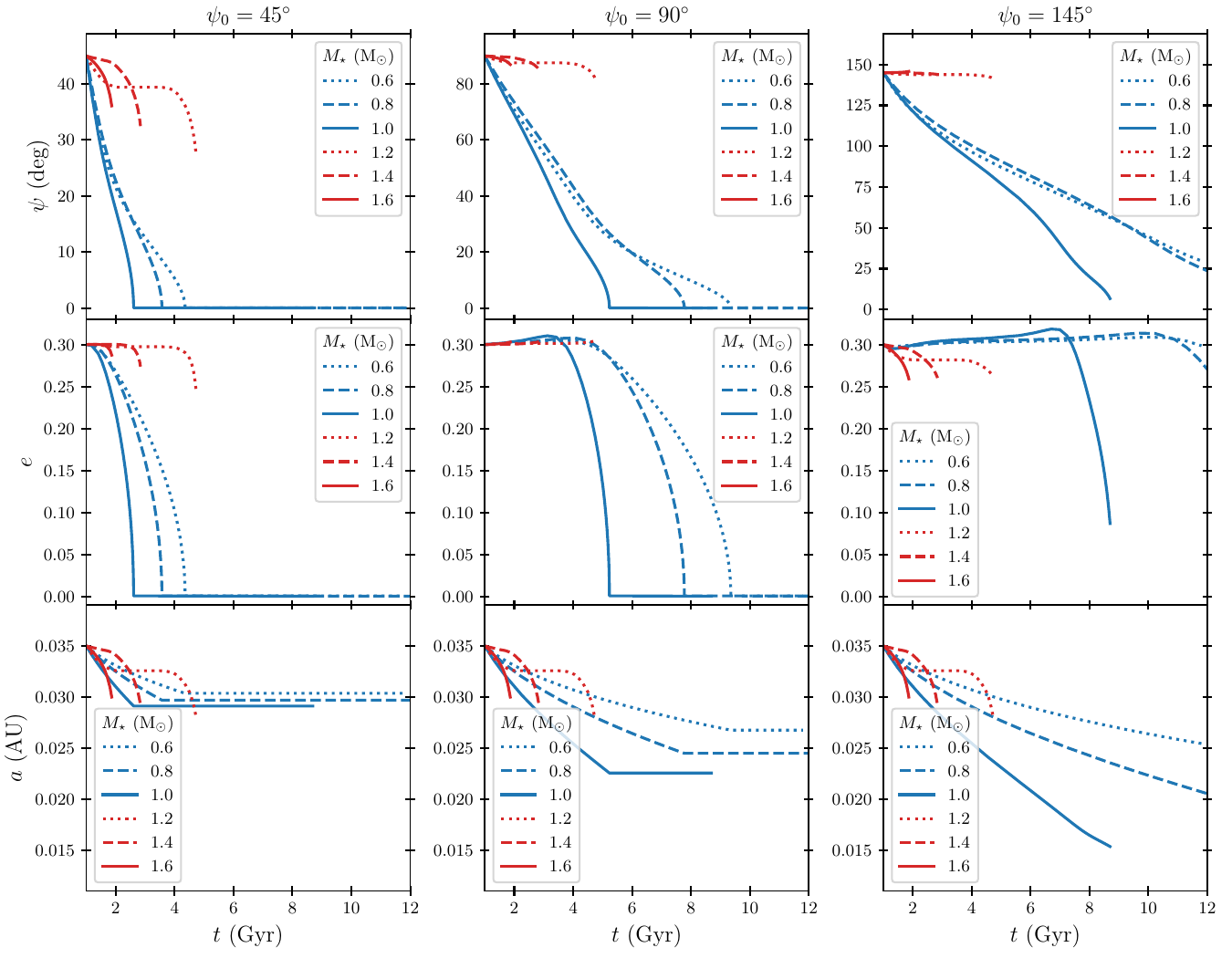}
\caption{
Same as Fig.~\ref{fig:orb_ev_circ}  except for initial eccentricities $e_0 = 0.3$. For such initial conditions, 
orbits can circularize around cool stars on the same timescale as the obliquity damps, but not hot stars.
The obliquity evolution is 
similar as for strictly circular orbits (compare with Fig.~\ref{fig:orb_ev_circ}).
\label{fig:orb_ev_lowe}
}
\end{figure*}

\begin{figure*}
\centering
\includegraphics[width=\linewidth]{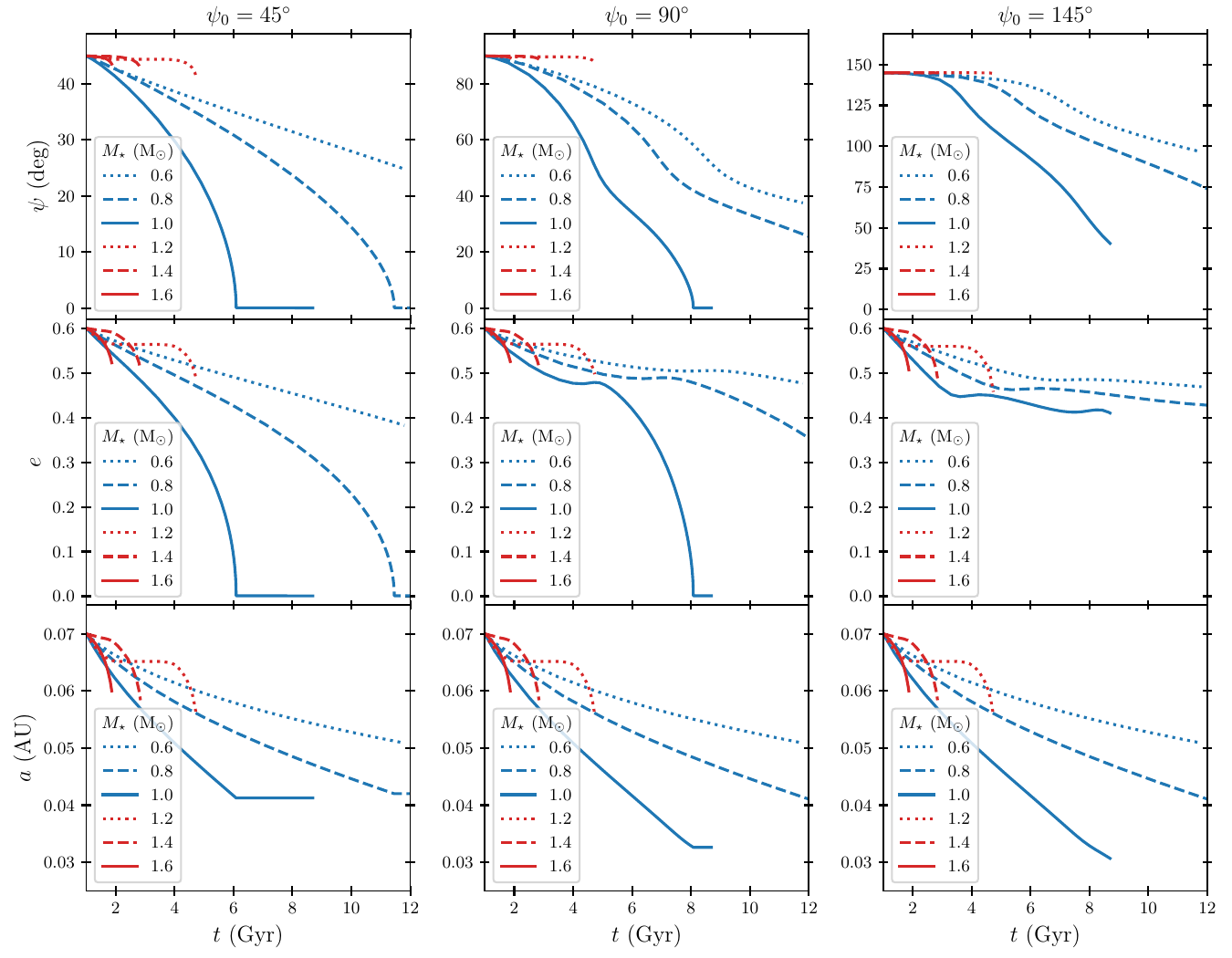}
\caption{
Same as Figs.~\ref{fig:orb_ev_circ} and \ref{fig:orb_ev_lowe}  except for initial conditions $\{e_0, a_0\} = \{0.6, 0.07 \, {\rm AU}\}$.  At these larger eccentricities, damping of $\psi$, $e$, and $a$ takes longer, but the equilibrium fixed point of $\psi, e = 0$ can still be reached within main-sequence lifetimes for cool stars whose obliquities are not too high. Hot star obliquities remain near their initial values regardless. 
\label{fig:orb_ev_highe}
}
\end{figure*}

\section{Results for Obliquity and Orbit Evolution}
\label{sec:Results}

We integrate equations~\eqref{eq:7}-\eqref{eq:10} for a planet of mass $\Mp = 1 \, {\rm M}_{\rm J}$ in resonant lock with a stellar gravity mode. We start all calculations at time $t_0 = 1 \, {\rm Gyr}$, and evolve $\psi$, $e$, and $a$ to  $t_{\rm max} = \min(t_{\rm MS}, 12 \, {\rm Gyr})$, where $t_{\rm MS}$ is the time when the stellar core 
stops burning hydrogen. 
Note that if and when both $\psi,e = 0$, resonance lock with our assumed $\{k,\ell,m\} = \{2,2,0\}$ mode cannot be maintained (e.g.~bottom left panel of Fig.~\ref{fig:ModeEx}), and $a$, $e$, and $\psi$ cease to change. In section \ref{subsec:break} and appendix~\ref{app:Nonlin_diss} we discuss more generally how resonance locks may actually break. Initial conditions vary over $a_0 = \{ 0.035, 0.07\}$ AU, $e_0 = \{ 0, 0.3, 0.6 \}$, and $\psi_0 = \{ 45^\circ, 90^\circ, 145^\circ\}$ (variables subscripted 0 are evaluated at time $t_0$).  We integrate our equations using the Python routine \texttt{solve\_ivp} from \texttt{scipy}, setting $\texttt{rtol} = 10^{-6}$ and $\texttt{atol} = 10^{-11}$, and evaluating $t_{\rm ev}^{-1}$ by linear interpolation (using \texttt{numpy} \texttt{interp}) of the data in the top panel of Fig.~\ref{fig:aveN_t}.

Stellar input parameters to equations~\eqref{eq:7}-\eqref{eq:10} are computed with \texttt{MESA} for stars of mass $\Ms = \{0.6, 0.8, 1.0, 1.2, 1.4, 1.6\} \,\Msun$; these parameters include $t_{\rm ev}$ (equations \ref{eq:N_ave}-\ref{eq:tev_brunt}), $\Rs(t)$, and
\be
\kg_\star(t) = \frac{8\pi}{3\Ms \Rs^2} \int_0^{\Rs} \rho r^4 \der r 
\label{eq:kgs}
\ee
where $\rho$ is the stellar mass density. 
We set the initial spin rate $\Omega_{\star 0} = 2\pi/(20 \ \der)$ for stars below the Kraft break (``cool'' or ``low-mass'', $\Ms < 1.2 \, \Msun$), and $\Omega_{\star 0} = 2\pi/(5 \ \der)$ for stars above (``hot'' or ``high-mass'', $\Ms \ge 1.2 \, \Msun$). These choices set $J_\star = \kg_\star \Ms \Rs^2 \Oms = \text{constant}$ (eq.~\ref{eq:8}), determining $\Oms(t)$ (which hardly changes in this simplified model). 
Our above choices for $\Om_{\star 0}$ match typical rotation rates of stars below and above the Kraft break (see e.g. Fig.~3 of \citealt{Albrecht+(2021)}).   For reference, as $\Ms$ increases from 0.6 to 1.6 $\Msun$ for our model stars, 
$\kappa_{\star 0}$ decreases from $\sim$0.15 to $\sim$0.04, $R_{\star 0}/a_0$ increases from $\sim$0.07 to $\sim$0.24 at $a_0 = 0.035$ AU, $\Omega_{\star 0}/\Omega_0$ increases from $\sim$$0.13$ to $\sim$$0.4$, 
and $J_{\rm orb, 0}/J_{\star 0}$ decreases from $\sim$$12.6$ to $\sim$$0.7$. For our low-mass stars, $J_{\rm orb}/J_{\star} > 1$ throughout the system evolution, a condition important for preferentially damping their obliquities 
(see equation \ref{eq:obl_damp_rate} and related discussion).

Figures \ref{fig:orb_ev_circ}, \ref{fig:orb_ev_lowe}, and \ref{fig:orb_ev_highe} present results for initial conditions $e_0 = 0$, 0.3, and 0.6, respectively. In all cases, obliquities $\psi$ damp more for low-mass stars than high-mass stars. For low-mass stars, obliquities can start as high as $90^\circ$-$145^\circ$ and drop to zero or nearly so within main-sequence lifetimes. Once the obliquity vanishes, the planet can no longer drive the $m=0$ mode that we are modeling; dissipation ceases and the planet stops migrating inward. For high-mass stars, obliquities hardly budge from their initial values. 
None of our planets is engulfed by its host star.

Larger eccentricities are seen to prolong obliquity damping and circularization times (compare Figs.~\ref{fig:orb_ev_circ}, \ref{fig:orb_ev_lowe}, and \ref{fig:orb_ev_highe}). A starting eccentricity of 0.3 damps to zero within main-sequence lifetimes for both low-mass and high-mass stars (Fig.~\ref{fig:orb_ev_lowe}), about as fast as the obliquity damps. At $e_0 = 0.6$, only the 1 and 0.8 $\Msun$ models circularize and achieve spin-orbit alignment within 12 Gyr for $\psi_0 \sim 45^\circ$ (Fig.~\ref{fig:orb_ev_highe}). 


\begin{figure}
\centering
\includegraphics[width=\linewidth]{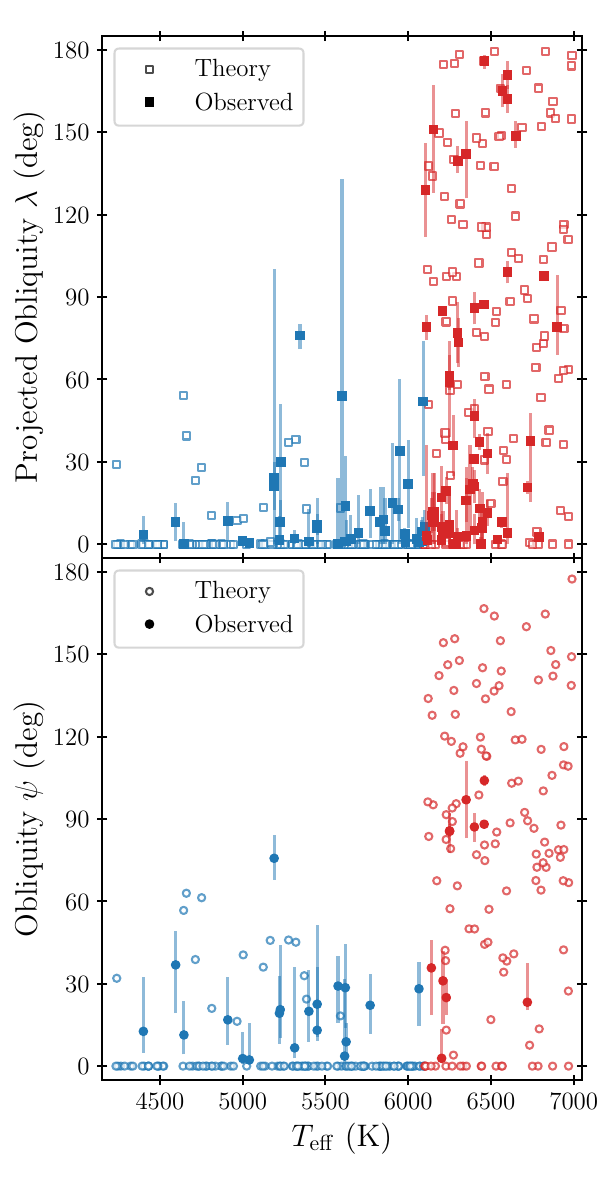}
\caption{
A population synthesis based on resonance locking (open symbols) compared against observations (filled symbols). Data for projected obliquities $\lambda$ (top) and 
obliquities $\psi$ (bottom) are taken from \cite{Albrecht+(2022)}, \cite{Rice+(2022), Rice+(2022b)}, \cite{Siegel+(2023)}, \cite{Espinoza-Retamal+(2023)}, \cite{Sedaghati+(2023)}, and \cite{Hu+(2024)}, for hot Jupiters ($M_{\rm p} \ge 0.3 \, {\rm M}_{\rm J}$, $a/\Rs \le 10$) above and below the Kraft break ($T_{\rm eff} \le 6100 \, {\rm K}$ in blue, $T_{\rm eff} > 6100 \, {\rm K}$ in red). In the model, 
obliquities are initially random (uniformly distributed in $\cos \psi_0$ between $-1$ and $1$), and subsequently damp by resonance locking over Gyr timescales. Damping is seen to be negligible for most hot stars (excepting those with $\psi_0 \lesssim 25^\circ$ which damp to zero), but significant for cool stars. Although roughly reproducing the break seen in the observations, the model fails to damp sufficiently cool star obliquities that start strongly retrograde; these evolve into the cloud of blue open symbols with obliquities between 
$10^\circ$ and $70^\circ$ sitting above the observations. 
\label{fig:Teff_comp}
 }
\end{figure}

\section{Discussion}
\label{sec:Disc}

We have shown that hot Jupiters locked in resonance with stellar g-modes can torque the stars into spin-orbit alignment and have their orbits circularized. The damping of obliquity and eccentricity is stronger for lower-mass stars because their g-mode frequencies evolve more on the main sequence. In the following subsections we delve deeper into the theory, and confront observations.

\subsection{Obliquity and Stellar Effective Temperature: \\Theory vs.~Observation}

We construct a simple population synthesis to compare against observations.
Host stellar masses $\Ms$ are sampled uniformly from the set $\{0.6, 0.8, 1.0, 1.2, 1.4, 1.6\} \ {\rm M}_\odot$. Around each star we consider a planet of mass $M_{\rm p} = 1 \, {\rm M}_{\rm J}$, initial semimajor axis $a_0 = 8 R_{\star 0}$, eccentricity $e_0 = 0$, and obliquity $\psi_0$ drawn randomly from a uniform distribution in $\cos \psi_0$ between $-1$ and 1 (for observational support of the latter distribution, see \citealt{Siegel+(2023)} and \citealt{DongForeman-Mackey(2023)}). Our equations of motion are then integrated from $t_0 = 1 \ {\rm Gyr}$ to $t_{\rm max}$ (either 12 Gyr or the main-sequence lifetime, whichever comes first), and the final values for $\psi$ extracted for comparison with observations. Since the observations typically trace stellar effective temperature $T_{\rm eff}$ instead of stellar mass, we map our set of 6 $\Ms$ values to the uniform intervals $\{[4200, 5000]$, $[5000, 5600]$, $[5600, 6100]$, $[6100, 6400]$, $[6400, 6700]$, $[6700, 7000]\} \ {\rm K}$, drawing randomly from each interval. Furthermore, since most measured obliquities are projected obliquities $\lambda$, we perform mock observations to convert 
obliquities into projected obliquities using 
\begin{equation}
    \tan \lambda \simeq \tan \psi \sin \phi_{\rm obs},
\end{equation} 
drawing $\phi_{\rm obs}$ randomly from a uniform distribution between 0 and $2\pi$ \citep[e.g.][]{FabryckyWinn(2009)}.


Figure~\ref{fig:Teff_comp} compares our modeled projected obliquities $\lambda$ with those observed (top panel; observations from \citealt{Rice+(2022)}). For completeness, we also show our modeled 
obliquities $\psi$ against the few  deprojected obliquities that are available (bottom panel; data from \citealt{Albrecht+(2022), Siegel+(2023)}). Broadly speaking, our model reproduces the dichotomy between hot oblique stars above the Kraft break, and cool aligned stars below. Obliquities of hot stars hardly change in our theory from their assumed initially isotropic distribution. Obliquities of cool stars are damped, all the way to zero 
if they start from 
$\psi_0 \lesssim 120^\circ$. Our model hot Jupiters have final semi-major axes $a \gtrsim 2.2 R_\star$ and thus avoid tidal disruption \citep[e.g.][]{Guillochon+(2011)} and stellar engulfment \citep[e.g.][]{BarkerOgilvie(2009), Winn+(2010), Dawson(2014)}. 

Fig.~\ref{fig:Teff_comp} also shows that our theory may over-predict the obliquities of the coolest (K-type) stars. If the initial obliquities are 
retrograde 
($\psi_0 \gtrsim 120^\circ$ for 0.6 $\Msun$ stars)
they evolve to become prograde but do not reach zero --- see the cloud of open symbols from our model at $T_{\rm eff} \lesssim 5100 \ {\rm K}$, and contrast with the filled symbols at low obliquity from observations. Other shortcomings of our population synthesis include our neglect of initially non-zero eccentricities (which would lengthen obliquity damping times and worsen the discrepancy between theory and observation for cool stars) and our neglect of a distribution of initial semi-major axes.

\subsection{Disabling Resonance Locks: $\psi_{\rm unlock}$ vs.~$a$} \label{subsec:break}

\begin{figure*}
\centering
\includegraphics[width=\linewidth]{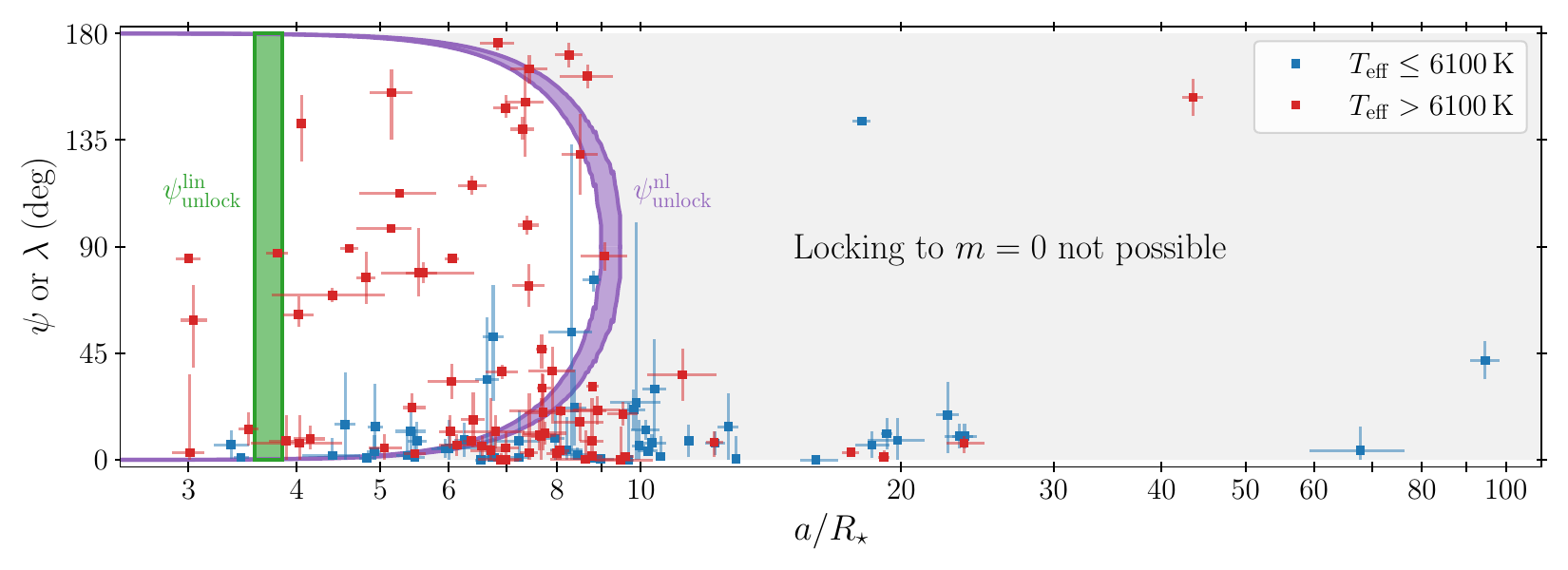}
\caption{Charting the limits of resonance locking. Computed curves are for an axisymmetric ($m=0$) g-mode of a $1\, {\rm M}_\odot$, 4-Gyr-old main-sequence star (see appendix \ref{app:Nonlin_diss}). The green bar ($\psi_{\rm unlock}^{\rm lin}$) marks a transition region dividing the space where resonance locking can operate (to the left) from where it cannot (to the right), computed under the assumption that tidal disturbances are linear and damp by radiative diffusion. In purple are plotted the analogous boundaries for $\psi_{\rm unlock}^{\rm nl}$, computed assuming tidal disturbances are limited in amplitude by non-linear mode-mode couplings, which spawn traveling waves that break and dissipate 
(see appendix \ref{app:Nonlin}). The non-linear model expands the regime where resonance locking is possible, out to $a/R_\star \approx 8$. Overlaid for context are observed projected obliquities $\lambda$ \citep{Albrecht+(2022), Rice+(2022), Rice+(2022b), Siegel+(2023), Espinoza-Retamal+(2023), Sedaghati+(2023), Hu+(2024)}. See section \ref{subsec:break} for how we might interpret these data.
\label{fig:psi_min}
}
\end{figure*}

As stellar obliquity decreases, tidal forcing of zonal g-modes weakens (see, e.g., the bottom left panel of Fig.~\ref{fig:ModeEx}).  There are obliquity boundaries $\psi_{\rm unlock}$ beyond which resonant locks fail. These boundaries depend on how tidal disturbances damp.

How tidally-driven modes damp is uncertain. We attempt two calculations, one using a ``linear'' damping rate for how waves lose energy to radiative diffusion when wave amplitudes are small, and another based on ``non-linear'' interactions that siphon energy away from tidally-driven modes to other stellar modes. Details are contained in appendix~\ref{app:Nonlin}.
Results for $\psi_{\rm unlock}$, computed for a 4-Gyr-old solar-mass star, 
are presented in Figure \ref{fig:psi_min}. According to the linear theory (green bar),  resonance locking can fully damp obliquities for $a/R_\star < 4$, but not for $a/R_\star > 5$.
Non-linear damping can extend the reach of resonance locks 
out to $a/R_\star \approx 8$-$10$ (purple curves).

Interestingly, as seen in Fig.~\ref{fig:psi_min}, most measured projected obliquities $\lambda$ of cool stars appear small at semi-major axes as far out as $a/R_\star \sim 100$ \citep[see also][]{Morgan+(2023)}. Our interpretation is that for stars beyond the resonance locking limit ($a/R_\star > 8$ according to the non-linear damping theory), small obliquities are primordial: Jupiter-mass planets accreted from disks aligned with stellar equatorial planes. This interpretation may also apply to the mostly small obliquities observed for hot stars with exo-Jupiters at large $a/R_\star \sim 20$ (with large-obliquity outliers tending to be in stellar binaries; \citealt{Rice+(2022b)}). By contrast, at $a/R_\star < 8$, hot stars are observed to have large obliquities, presumably the result of whatever process created hot Jupiters (e.g.~planet-planet scatterings / high-eccentricity migration). Hot star obliquities hardly damp from resonance locking because the g-mode frequencies of hot stars, which lack radiative cores, change by $\lesssim 10\%$ over most of their main sequence lives (sections \ref{sec:GModeEv} and \ref{sec:Results}). If hot Jupiters form around cool stars the same way they do around hot stars, they would be similarly initially misaligned. We have shown that cool star misalignments can be subsequently erased in resonance lock, as their radiative cores become more strongly stratified and g-mode frequencies increase.

We reiterate that the model curves for $\psi_{\rm unlock}$ in Fig.~\ref{fig:psi_min} are specific to a $1 \,\Msun$ star, and uncertain as they depend on how tidally-excited modes damp. The damping mechanism is not merely a technical detail; its nature determines whether resonance locking is possible at all. \cite{MaFuller(2021)} did not expect hot Jupiters to resonantly lock to their host stars because giant planet perturbations may be sufficiently strong to trigger a parametric instability that spawns `child' modes. When child modes take the form of standing waves and are fully excited, their energy dissipation rates are insensitive to distance from resonance ($k\Omega - \sigma_\alpha$), and resonance locking is defeated \citep{EssickWeinberg(2016)}. In appendix \ref{app:Nonlin} we review this argument  and rebut it with the possibility that  child modes may be so strong as to overturn, becoming traveling waves. Traveling waves dissipate qualitatively differently from standing waves, and we offer some exploratory calculations in the appendix showing how the resonant response of the star may be restored. The purple `non-linear' curve for $\psi^{\rm nl}_{\rm unlock}$ in Fig.~\ref{fig:psi_min} is computed under this tentative traveling-child picture.

\subsection{Stellar Ages and Resonance Locking Duration}\label{subsec:t0}

Our calculations of resonance locking start at $t_0 = 1 \ {\rm Gyr}$, when our higher-mass stars have only $\sim$1--3 Gyr left before they ascend the subgiant branch. We have checked that the calculations shown  Figs.~\ref{fig:orb_ev_circ}-\ref{fig:orb_ev_highe} are robust to changes in $t_0$ down to 0.1 Gyr. Starting times of 0.1--1 Gyr appear consistent with high-eccentricity migration, which takes of order such times to form hot Jupiters on initially misaligned orbits \citep[e.g.][]{DawsonJohnson(2018),Wu(2018),Vick+(2019)}. Similarly long gestation times are indicated by the observation that high-mass, misaligned hot Jupiter hosts have higher Galactic velocity dispersions and are therefore older than aligned hosts \citep{HamerSchlaufman(2022)}.

Late formation of misaligned hot Jupiters bypasses resonance locks that occur during the pre-main sequence \citep{ZanazziWu(2021)}. Pre-main sequence locks would predict little difference in tidal evolution with stellar mass, as young stars have similar internal structures (mostly convective) irrespective of mass \citep[e.g.][]{Henyey+(1955)}.


\subsection{Obliquity vs.~Eccentricity}

We have found that resonance locking damps eccentricity, more so for lower mass stars. Compared to obliquities, however, eccentricities damp more slowly  (Figs.~\ref{fig:orb_ev_lowe} and \ref{fig:orb_ev_highe}), suggesting that some hot Jupiters should be found at low $\lambda$ and high $e$ --- in conflict with the observed absence of such objects around cool stars (Fig.~\ref{fig:ecc_data}).
Tidal dissipation in the planet, rather than the star, may be more effective at damping eccentricity \citep[e.g.][]{Rice+(2022)}.

\subsection{Obliquity vs.~Planet Mass}

We have focused in this work on Jupiter-mass planets as these have some of the most reliable obliquity measurements. Lower planet masses decrease obliquity damping rates (approximately linearly through the term $J_{\rm orb} \propto \Mp$ in equation~\ref{eq:10}). 
Accordingly, we would not expect stars hosting hot sub-Neptunes to exhibit the same obliquity trends shown by stars hosting hot Jupiters. Observations appear to support this expectation \citep[e.g.][]{Herbrard+(2011),Albrecht+(2022)}, although \cite{Louden+(2021)} found from rotational broadening measurements that cool host stars of sub-Neptunes may be more aligned than hot host stars.



Conversely, more massive planets should damp stellar obliquities faster. 
\cite{Herbrard+(2011)} and 
\cite{Albrecht+(2022)} have noted that ``moderately'' hot stars ($6250 \, {\rm K} \lesssim T_{\rm eff} \lesssim 7000 \, {\rm K}$; $1.3 \, {\rm M}_\odot \lesssim \Ms \lesssim 1.6 \, {\rm M}_\odot$), when orbited by brown dwarfs ($\Mp \gtrsim 10 \, {\rm M}_{\rm J}$), display small obliquities (see e.g.~\citealt{Albrecht+(2022)} Figure 9, middle panel).


\subsection{Stellar Rotation}

We draw stellar rotation rates from observations \citep[e.g.][Fig.~3]{Albrecht+(2021)}: 20-day rotation periods for low-mass stars, and 5-day rotation periods for high-mass stars. The lower angular momenta of low-mass stars increases their rate of obliquity damping --- see the term $J_{\rm orb}/J_\star$ in eq.~\ref{eq:obl_damp_rate}  --- abetting the effects of core hydrogen burning (smaller $t_{\rm ev}$).
As stars spin down from magnetic braking, they should align faster. It may be possible to test, for a given stellar mass, whether slower rotating hosts are more aligned.

\subsection{Non-Axisymmetric Modes}\label{subsec:nonaxi}

We have focussed in this work on the $m=0$ axisymmetric stellar g-mode and how it damps obliquity. 
In appendix \ref{app:EL_tide_loss} we show that obliquity damps for a g-mode for all $|m| \leq \ell$, when forced into resonance lock by the $\{k, \ell\} = \{2,2\}$ component of the tidal potential of a  
circular planet ($e=0$)
with $J_\ro > J_\star$. The obliquity damps as (eq.~\ref{eq:obl_damp_full})
\begin{equation} \label{eq:psidamp}
    \frac{\der \psi}{\der t} = - \left[  \tau_m \left( \frac{J_\ro}{J_\star} + \cos \psi \right) - m \sin \psi \right] \frac{1}{4 \eta_\ag t_{\rm ev}}.
\end{equation}
The parameter $0<\eta_\ag\le3/2$ for typical hot Jupiter systems (eq.~\ref{eq:eta}); if the planet mass is too large  ($\Mp \gtrsim {\rm few} \ \Mjup$), then $\eta_\ag \le 0$ and resonance locking is not possible.
For $m=0$, the torque coefficient $\tau_0 = 2/\sin \psi$, while for $m=\pm 1$ and $m=\pm 2$,
\begin{equation}
\tau_{\pm 1} = \frac{2 \mp \cos \psi}{\sin \psi}, \hspace{5mm} \tau_{\pm 2} = \frac{2 \sin \psi}{1 \pm \cos \psi} .
\end{equation}
These torque coefficients are always positive, and for $J_\ro~>~J_\star$, they lead to the first term in (\ref{eq:psidamp}) dominating the second term, giving $\der \psi/\der t < 0$ for all $\psi$.  

Calculating the detailed time history of $\psi$ for $m\neq 0$ is left for future work. Here we offer some comments and speculation. The likelihood of locking onto a mode of given $m$ should be proportional to the amplitude of the tidal potential $|U^2_{2m2}|$ (eq.~\ref{eq:Uklm}).
Whereas polar planets ($\psi  \approx \pi/2$) are favored to lock onto zonal g-modes ($m=0$), 
prograde obliquities ($\psi < \pi/2$) will likely lock onto prograde modes ($m>0$), and retrograde obliquities ($\psi > \pi/2$) onto retrograde modes ($m<0$). 
Because prograde locks have weaker alignment torques ($\tau_1, \tau_2 < \tau_0$ when $\psi<\pi/2$) and spin up the host star ($\dot J_\star \propto m > 0$, eq.~\ref{eq:dJsdt_full}), they will damp prograde obliquities more slowly than zonal locks. 
Sectoral ($m=2$) locks may help explain the rapid rotation rates of hot Jupiter hosts \citep{Penev+(2018), MaFuller(2021)}, 
but also damp semi-major axes and obliquities at comparable rates (similar to equilibrium tides), thus raising the possibility of engulfment before alignment.
Relative to zonal locks, retrograde locks may damp initially retrograde obliquities more slowly, but once the obliquity swings to prograde, damping may be faster because alignment torques are stronger ($\tau_{-2}, \tau_{-1} > \tau_0$ when $\psi<\pi/2$), and because retrograde locks spin down the star ($\dot J_\star < 0$).





\section{Summary}
\label{sec:Conc}

A planet can tidally force a stellar gravity mode (g mode), with consequences for the planet's orbit and the star's spin. Structural changes inside the star over the course of its evolution can change the companion's orbit so as to maintain a commensurability between the orbital frequency and the g-mode oscillation frequency. Such resonance locking can alter orbital semi-major axes and eccentricities, and stellar rotation rates \citep[e.g.][]{ZanazziWu(2021), MaFuller(2021)}. The direction of the stellar spin axis relative to the planet's orbit normal can also change in resonance lock, as we have established in this paper.

Stars that host hot Jupiters are known to have spin axes misaligned from their orbit normals. Large stellar obliquities are thought to be a relic of the gravitational scatterings and long-term dynamical perturbations that originally delivered hot Jupiters onto their close-in orbits \citep[e.g.][]{DawsonJohnson(2018)}. Primordial spin-orbit misalignments appear to have been preserved for host stars with effective temperatures $T_{\rm eff} \gtrsim 6100$ K, but may have damped away for cooler stars with near-circular planets \citep{Winn+(2010), Rice+(2022)}. We have shown how obliquity damping goes hand-in-hand with semi-major axis and eccentricity damping when star and planet are resonantly locked. While on the main sequence, cooler stars experience much more damping than hotter stars. The dependence on $T_{\rm eff}$ arises because stellar g-modes are sustained only in stably stratified radiative zones, and cool stars have radiative cores whose g-mode (Brunt-V\"ais\"al\"a) frequencies increase substantially from core hydrogen burning, thereby driving significant tidal evolution. Hotter stars lack such cores. Thus resonance locking can explain how cooler stars are aligned with hot Jupiters while hotter stars are not.

There are many areas where our work can be improved.
As a simplification, we assumed throughout most of our study that the planet locks to an axisymmetric (zonal) gravity mode. 
The obliquity damps whether the planet locks to an axisymmetric or non-axisymmetric mode, but the efficiency of damping will vary with the azimuthal wavenumber $m$, possibly significantly. 
More importantly, our work rests on an uncertain foundation. Resonance locking is not possible if tidally-excited stellar oscillations decay into standing-wave `child' modes whose energy dissipation does not depend on proximity to resonance \citep{EssickWeinberg(2016),MaFuller(2021)}. We have proposed a way out, whereby child standing waves break to become traveling waves which damp qualitatively differently  (appendix \ref{app:Nonlin}). Our preliminary calculations in this regard appear promising but need further testing and development, probably from hydrodynamic simulations.




\vspace{0.12in}
\noindent We thank 
Simon Albrecht, Linhao Ma, Diego Munoz, Malena Rice, Joshua Winn, and Yanqin Wu for discussions. An anonymous referee provided an insightful report and motivated our ongoing work on tidal damping (appendix~\ref{app:Nonlin}). Financial support was provided by a 51 Pegasi b Heising-Simons Fellowship awarded to JJZ. JWD was supported by the Natural Sciences and Engineering Research Council of Canada (NSERC) [funding reference \#CITA 490888-16]. EC acknowledges support from a Simons Investigator award and NSF AST grant 2205500.

\vspace{5mm}

\software{astropy \citep{Astropy_1,Astropy_2},  
          GYRE \citep{TownsendTeitler(2013), TownsendZweibel(2018), GoldsteinTownsend(2020)}
          MESA \citep{Paxton+(2011),Paxton+(2013),Paxton+(2015),Paxton+(2018),Paxton+(2019),Jermyn+(2023)}, 
          numpy \citep{numpy_cite},
          pandas \citep{pandas_cite},
          PyVista \citep{pyvista},
          scipy \citep{scipy_cite}
          }

\appendix

\section{Energy and Angular Momentum Transfer from an Inclined, Eccentric Planet to a Spinning Star}
\label{app:EL_tide_loss}

This appendix considers the general case of an eccentric planet inclined to the star's equatorial plane, and calculates the energy and angular momentum exchange rate between the orbit of the planet, and a general stellar oscillation.  We also calculate the orbital evolution which results from this angular momentum transfer.

\subsection{Tidal Potential}\label{subsec:app_tide_pot}

\begin{figure*}
\centering
\includegraphics[width=0.6\linewidth]{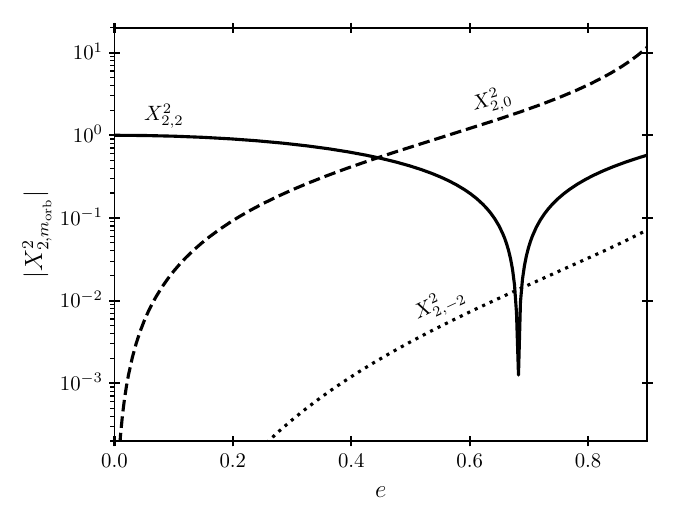}
\caption{
Plotting the magnitude of the Hansen coefficients (eq.~\ref{eq:Xklm}), for the $\{k, \ell\} = \{2,2\}$ component of the tidal potential.  The $|X^2_{2,2}|$ coefficient dip at $e \sim 0.7$ occurs because $X^2_{2,2}$ changes sign.  The $|X^2_{2,2}|$ coefficient is larger than $|X^2_{2,0}|$ at smaller eccentricities, until $e \gtrsim 0.4-0.5$.  We always have $|X^2_{2,-2}| \ll |X^2_{2,0}|$.
\label{fig:hansen}
}
\end{figure*}

In a spherical coordinate system $(r, \theta_\ro, \vphi_\ro)$ centered on the star, with the $z_\ro$-axis parallel to the planet's orbit normal and the $x_\ro$-axis pointing towards pericenter, the tidal potential is given by \citep[e.g.][]{ZanazziWu(2021)}
\begin{align}
        &U = - \frac{G \Mp}{a} \sum_{\ell=2}^\infty \sum_{m_\ro = -\ell}^{\ell} \left( \frac{r}{a} \right)^{\ell} W_{\ell m_\ro} \sum_{k=-\infty}^{\infty} X_{\ell m_\ro}^k (e) \e^{-\im k \Om t} Y_{\ell m_\ro}(\theta_\ro, \vphi_\ro).
        \label{eq:U_planet}
\end{align}
Here $k$ is the orbital harmonic, $\ell$ is the angular degree, $m_\ro$ is the azimuthal number of the tidal potential in a coordinate system aligned with the planet's orbital plane, $W_{\ell m_\ro}$ is given by equation~(24) of \cite{PressTeukolsky(1977)}, and
\begin{equation}
    X_{\ell m_\ro}^k(e) = \frac{1}{\pi} \int_0^\pi \der E \frac{\cos[k(E - e \sin E) - m_\ro f]}{(1-e \cos E)^\ell},
    \label{eq:Xklm}
\end{equation}
are Hansen coefficients, with $f$ the true anomaly.  Notice the eccentricity $e$ and exponential $\e$ differ typographically.  For $\ell=2$, the non-zero $W_{\ell m_\ro}$ values are $W_{2,\pm 2} = \sqrt{3\pi/10}$ and $W_{20} = -\sqrt{\pi/5}$.  For $e \ll 1$ \citep{Weinberg+(2012)}
\begin{equation}
    X_{\ell m_\ro}^k(e) = \dg_{km_\ro} 
    + \frac{1}{2}e \left[(\ell+1-2m_\ro) \dg_{k,m_\ro-1} + (\ell+1+2m_\ro)\dg_{k,m_\ro+1} \right] + \cO(e^2).
\end{equation}
Figure~\ref{fig:hansen} plots the coefficients $|X^2_{2,m_\ro}|$ relevant for our obliquity integrations.
For $e \lesssim 0.4$, the $m_\ro = 2$ Hansen coefficient plotted in Figure~\ref{fig:ModeEx} has the largest magnitude. At larger $e$, the $m_\ro = 0$ Hansen coefficient dominates.

We transform $(x_\ro, y_\ro, z_\ro)$ to a coordinate system with the $z$-axis parallel to the stellar spin axis $\hjs$, and the $y$-axis parallel to $\hjo \btimes \hjs$ (the orbit's node), co-rotating with the stellar spin.  After rotating $Y_{\ell m_\ro}(\theta_\ro, \vphi_\ro)$ by the Euler angles ($\ag_E, \bg_E, \cg_E) = (-\Om_\star t, \psi, \om_p + \pi/2)$, where $\om_p$ is the argument of pericenter, we obtain
\begin{equation}
    Y_{\ell m_\ro}(\theta_\ro, \vphi_\ro) = \sum_{m=-\ell}^\ell D^\ell_{m m_\ro}(\ag_E, \bg_E, \cg_E) Y_{\ell m}(\theta, \vphi).
    \label{eq:Ylm_exp}
\end{equation}
Here, $m$ is the azimuthal number in the star's rotating frame, and $D^\ell_{m m_\ro}(\ag_E, \bg_E, \cg_E)$ are Wigner $D$-functions using the $z$-$y$-$z$ convention of \cite{Varshalovich+(1988)}.  Inserting equation~\eqref{eq:Ylm_exp} into~\eqref{eq:U_planet}, we obtain the tidal potential in the star's rotating frame:
\begin{equation}
    U(t,r,\theta,\vphi) = - \frac{G M_\star}{R_\star} \sum_{k=-\infty}^{\infty} \sum_{\ell=2}^\infty \sum_{m=-\ell}^\ell \sum_{m_\ro=-\ell}^\ell \left( \frac{r}{R_\star} \right)^{\ell} U_{\ell m m_\ro}^k Y_{\ell m}(\theta,\vphi) \e^{-\im \om_{km} t},
    \label{eq:Ust}
\end{equation}
where
\begin{align}
    &U_{\ell m m_\ro}^k = \frac{\Mp}{M_\star} \left( \frac{R_\star}{a} \right)^{\ell+1} Z_{\ell m m_\ro}^k,
    \label{eq:Uklm}\\
    &Z_{\ell m m_\ro}^k(e,\psi, \om_p) = W_{\ell m_\ro} X_{\ell m_\ro}^k(e) d_{m m_\ro}^\ell(\psi) \e^{-\im m_\ro (\om_p+\pi/2)},
    \label{eq:Zklm}\\
    &\om_{km} = k \Om - m \Oms.
    \label{eq:omkm}
\end{align}
Note that $U$ is real.

The Wigner $D$-function is related to the Wigner small-$d$ functions via $D_{mm_\ro}^\ell(\alpha_E, \beta_E, \gamma_E) = \e^{-\im m \ag_E} d_{m m_\ro}^\ell (\beta_E) \e^{-\im m_\ro \gamma_E}$.  Fixing $\ell=2$, because $W_{\ell m_\ro}$ is non-zero only for $m_\ro = \pm 2, 0$, the only contributing $d_{m m_\ro}^\ell(\beta_E)$ values are
\begin{align}
    &d^2_{\pm 2, 2} = \frac{1}{4}(1 \pm \cos \bg_E)^2, &
    &d^2_{\pm 1, 2} = \frac{1}{2} \sin \bg_E (1 \pm \cos \bg_E), &
    &d^2_{0,2} = \sqrt{ \frac{3}{8} } \sin^2 \bg_E, \nonumber \\
    &d^2_{\pm 2, 0} = \sqrt{ \frac{3}{8} } \sin^2 \bg_E, &
    &d^2_{\pm 1, 0} = \mp \sqrt{ \frac{3}{2} } \sin \bg_E \cos \bg_E, &
    &d^2_{0,0} = \frac{1}{2}(3 \cos^2 \bg_E - 1), \nonumber \\
    &d^2_{\pm 2, -2} = \frac{1}{4} (1 \mp \cos \bg_E)^2, & 
    &d^2_{\pm 1, -2} = - \frac{1}{2} \sin \bg_E (1 \mp \cos \bg_E), &
    &d^2_{0,-2} = \sqrt{ \frac{3}{8} } \sin^2 \bg_E.
    \label{eq:Wigner_d}
\end{align}

\subsection{Mode Amplitude}\label{subsec:app_mode_amp}

If we decompose both the displacement $\bxi(t,\br)$ and its time derivative as sums over g-mode oscillations $\bxi_\ag(\br)$ with eigenfrequencies $\om_\ag$ and dimensionless amplitudes $c_\ag(t)$, e.g., 
\begin{equation}
    \bxi(t, \br) = \sum_\ag c_\ag(t) \bxi_\ag(\br),
    \label{eq:ca_def}
\end{equation}
the mode amplitude evolves as \citep[e.g.][]{Schenk+(2001),LaiWu(2006)}
\begin{equation}
    \dot c_\ag + (\im \om_\ag + \cg_\ag) c_\ag = \im \om_\ag \sum_{k=-\infty}^{\infty} \sum_{\ell=2}^\infty \sum_{m=-\ell}^\ell \sum_{m_\ro=-\ell}^\ell U_{\ell m m_\ro}^k \left( \frac{E_\star}{\eps_\ag} \right) I_{\ag \ell m} \e^{-\im \om_{km} t},
    \label{eq:dcdt_tot}
\end{equation}
where $E_\star = G \Ms^2/\Rs$ is the host's binding energy,
\begin{equation}
    \eps_\ag = 2 \om_\ag^2 \int \bxi_\ag^* \bcdot \bxi_\ag \rho \der V
\end{equation}
is the mode energy (added to its complex conjugate), while the overlap integral is \citep[e.g.][]{Weinberg+(2012)}
\begin{equation}
    I_{\ag \ell m} = \frac{1}{M R^\ell} \int_0^{\Rs} r^{\ell+2} \dg \rho_{\ag \ell m}^* \der r .
\end{equation}
Equation~\eqref{eq:dcdt_tot} has the inhomogeneous solution
\begin{equation}
    c_\ag = \sum_{k=-\infty}^{\infty} \sum_{\ell=2}^\infty \sum_{m=-\ell}^\ell \sum_{m_\ro=-\ell}^\ell c_{\ag \ell m m_\ro}^k \e^{-\im \om_{km} t}
\end{equation}
in the rotating frame of the star, which transforms to
\begin{equation}
    c_{\ag,{\rm inert}} = \sum_{k=-\infty}^{\infty} \sum_{\ell=2}^\infty \sum_{m=-\ell}^\ell \sum_{m_\ro=-\ell}^\ell c_{\ag \ell m m_\ro}^k \e^{-\im k \Om t}
    \label{eq:ca_inert}
\end{equation}
in the inertial frame, where
\begin{equation}
    c_{\ag \ell m m_\ro}^k = \frac{\om_\ag}{\om_\ag - \om_{km} - \im \cg_\ag} U_{\ell m m_\ro}^k \left( \frac{E_\star}{\eps_\ag} \right) I_{\ag \ell m}.
    \label{eq:ca_force}
\end{equation}

Notice that when $\om_{km} \approx \om_\ag$, a single oscillation $\ag$ forced by the tidal component $\{k, \ell, m\}$ has an amplitude much larger than other oscillations.  For this reason, we will consider a single oscillation $\ag$ forced by a fixed component of the tidal potential $\{k, \ell, m\}$, and drop subscripts $\{\ag, k, \ell\}$ on all quantities for the remainder of appendix~\ref{app:EL_tide_loss} (but keep $\{m, m_\ro\}$). Also, we will work solely in the inertial frame, and drop the subscript ``inert'' on equation~\eqref{eq:ca_inert}.

\subsection{Energy and Angular Momentum Transfer}

Equation~\eqref{eq:Ust} shows each tidal component $\{k, \ell, m\}$ is forced by multiple orbital azimuthal numbers $m_\ro$, each of which differs in phase by $m_\ro \om_p$.  Because we expect short-range forces to cause $\om_p$ to precess over a timescale much shorter than $\sim$1 Gyr \citep[e.g.][]{Liu+(2015)}, we will average over the planet's argument of pericenter, on top of azimuthal and time averages.  Every orbital azimuthal number $m_\ro$ from an eccentric planet's tidal potential, forcing a g-mode with azimuthal number $m$, is included in our calculation.

\subsubsection{Energy Transfer}

The energy transferred from the planet's orbit to the star is \citep[e.g.][]{Lai(2012)}:
\begin{equation}
    \dot E = \int \dot \bxi^* \bcdot \bdel (-U) \rho \der V + \int \dot \bxi \bcdot \bdel (-U^*) \rho \der V = -2 \int \Re \left( \frac{\pd}{\pd t} \dg \rho^* \ U \right) \der V.
\end{equation}
Here, $\Re(A)$ denotes the real part of $A$, and we have added $\bxi$ to its complex conjugate \citep[e.g.][]{Schenk+(2001), Burkart+(2014)}.  
This latter procedure is for bookkeeping the complex addends in expansion~\eqref{eq:ca_def}, since we focus on modes with eigenfrequencies $\om_\ag > 0$.
Inserting equations~\eqref{eq:Ust} and~\eqref{eq:ca_inert} after averaging over time,
\begin{equation}
    \dot E_m = 2 \Re \left[ \im  k \Om \left( \frac{G M^2}{\Rs} \right) I_m \sum_{m_\ro=-2}^2 c_{m m_\ro}^* U_{m m_\ro} \right].
\end{equation}
Also averaging over $\om_p$, this reduces to 
\begin{equation}
    \langle \dot E_m \rangle = \frac{1}{2\pi} \int_0^{2\pi} \dot E_m \ \der \omega_p = 2 \frac{G \Ms^2}{\Rs} k \Om  \frac{\om \cg }{(\om - \om_m)^2 + \cg^2}  |I_m|^2 \left( \frac{E_\star}{\eps} \right) \sum_{m_\ro=-2}^2 |U_{m m_\ro}|^2.
    \label{eq:dotEaklm}
\end{equation}

\subsubsection{Angular Momentum Transfer ($z$-direction)}

The torque in the $z$-direction is \citep[e.g.][]{Lai(2012)}
\begin{equation}
    T_z = \int \dg \rho^* L_z (-U) \der V + \int \dg \rho L_z (-U^*) \der V = 2 \int \Re \left[ \rho^* L_z (-U) \right] \der V,
    \label{eq:Tz}
\end{equation}
where
\begin{equation}
    L_z = \frac{\pd}{\pd \vphi}
\end{equation}
is the $z$-component of the angular momentum operator.  Inserting equation~\eqref{eq:Ust} and averaging over one orbital period, we find
\begin{equation}
    T_{z,m} = 2 \Re \left[ \im m \left( \frac{G M_\star^2}{\Rs} \right) I_m  \sum_{m_\ro = -2}^2 c_{m m_\ro}^* U_{m m_\ro} \right].
\end{equation}
Also averaging over $\om_p$, this reduces to
\begin{equation}
    \langle T_{z,m} \rangle = \frac{1}{2\pi} \int_0^{2\pi} T_{z,m} \ \der \omega_p  = 2 \frac{G M_\star^2}{\Rs} m  \frac{\om \cg}{(\om - \om_m)^2 + \cg^2} |I_m|^2 \left( \frac{E_\star}{\eps} \right)  \sum_{m_\ro=-2}^2 |U_{m m_\ro}|^2.
    \label{eq:Tz_ave}
\end{equation}
Notice that
\begin{equation}
    \langle T_{z,m} \rangle = \frac{m}{k \Om} \langle \dot E_m \rangle,
    \label{eq:Tzaklm}
\end{equation}
as expected \citep[e.g.][]{FriedmanSchutz(1978a), FriedmanSchutz(1978b)}.  

\subsubsection{Angular Momentum Transfer ($x$ and $y$ directions)}

The torque in the $x$ and $y$ directions are \citep[e.g.][]{Lai(2012)}
\begin{equation}
    T_x = 2 \int \Re \left[ \dg \rho^* L_x (-U) \right] \der V, \hspace{5mm}
    T_y = 2 \int \Re \left[ \dg \rho^* L_y (-U) \right] \der V,
\end{equation}
where
\begin{equation}
    L_x = -\sin \vphi \frac{\pd}{\pd \theta} - \cot \theta \cos \vphi \frac{\pd}{\pd \vphi}, \hspace{5mm} L_y = \cos\vphi \frac{\pd}{\pd \theta} - \cot\theta \sin\vphi \frac{\pd}{\pd \vphi},
\end{equation}
are the angular momentum operators in the $x$ and $y$ directions.  Re-writing these in terms of the raising and lowering operators in quantum mechanics:
\begin{equation}
    L_x = \frac{1}{2}(L_+ + L_-), \hspace{5mm} L_y = -\frac{1}{2} \im (L_+ - L_-),
\end{equation}
where
\begin{equation}
    L_{\pm} = \im \e^{\pm i \vphi} \left( \pm \frac{\pd}{\pd \theta} + \im \cot\theta \frac{\pd}{\pd \vphi} \right),
\end{equation}
and using the well-known relation with spherical harmonics
\begin{equation}
    L_{\pm} Y_{m} = \im \Lambda_m^{\pm} Y_{m \pm 1},
\end{equation}
where
\begin{equation}
    \Lambda_{m}^{\pm} = \sqrt{\ell(\ell+1) - m(m\pm 1)},
\end{equation}
inserting equation~\eqref{eq:Ust} into $T_x$ and $T_y$ gives
\begin{align}
    T_{x,m} &= \frac{G \Ms^2}{\Rs} \Re \left\{ \im  I_m \sum_{m_\ro = -2}^2 c_{m m_\ro}^*  \left[ \Lambda_m^+ U_{(m+1)m_\ro} + \Lambda^-_m U_{(m-1)m_\ro} \right] \right\}, \\
    T_{y,m} &=  \frac{G \Ms^2}{\Rs}  \Re \left\{ - I_m \sum_{m_\ro = -2}^2 c_{m m_\ro}^*  \left[ \Lambda_m^+ U_{(m+1)m_\ro} - \Lambda^-_m U_{(m-1)m_\ro} \right] \right\}.
\end{align}
Averaging $T_{x,m}$ over $\om_p$, we may write
\begin{align}
    \langle T_{x,m} \rangle &= \frac{1}{2\pi} \int_0^{2\pi} T_{x,m} \ \der \omega_p =   \frac{1}{2m} \left( \tau_m^+ + \tau_m^- \right) \langle T_{z,m} \rangle = \frac{1}{2 k \Om} \left( \tau_m^+ + \tau_m^- \right) \langle \dot E_m \rangle, \label{eq:Txaklm}
\end{align}
where the real coefficients
\begin{equation}
    \tau_m^\pm = \left[ \sum_{m_\ro=-2}^2 \Lambda_m^\pm Z_{m m_\ro}^* Z_{(m \pm 1) m_\ro} \right] \Bigg/ \left[  \sum_{m_\ro=-2}^2 |Z_{m m_\ro}|^2  \right]. \label{eq:tau_klm}
\end{equation}
We ignore $T_y$, as it causes precession of the spin axis about the orbit normal.

\subsection{Orbital Evolution for Arbitrary Resonance}

Now, consider an oscillation frequency in the inertial frame  $\sg$ (mode index $\ag$, and angular/azumuthal number $\ell$, $m$ suppressed).  Resonance locking occurs when an integer harmonic $k$ of the orbital frequency satisfies
\be
k \Om \simeq \sg.
\ee
Differentiating with time and re-arranging, one can show in the absence of other forces (e.g. magnetic breaking, contraction, additional planet/star tides), that \citep{Fuller(2017)}
\begin{equation}
    t_{\rm ev}^{-1} \equiv \frac{\dot \sg}{\sg} = \eta \frac{\dot E_\ro}{E_\ro},
    \label{eq:tev_de_appendix}
\end{equation}
where
\be
\eta = \frac{3}{2} - \frac{m^2 B}{2 k^2 \sqrt{1-e^2}} \left( \frac{\Mp a^2}{\kg_\star M_\star \Rs^2} \right),
\label{eq:eta}
\ee
and (assuming $\Oms \ll \sg$)
\be
B = \frac{1}{m} \frac{\pd \sg}{\pd \Oms} \simeq 1 - \frac{1}{\ell(\ell+1)}.
\ee
Inserting the averaged equations~\eqref{eq:Tzaklm} and~\eqref{eq:Txaklm} into~\eqref{eq:1}-\eqref{eq:4}, and re-arranging $\dot J_\ro$ and $\dot E_\ro$ to get $\der e/\der t$, we find
\begin{align}
    \frac{1}{a} \frac{\der a}{\der t} = &- \frac{1}{\eta t_{\rm ev}}, \label{eq:a_damp_full} \\
    \frac{\der J_\star}{\der t} = & \ \frac{m}{2 k \sqrt{1-e^2}} \frac{J_\ro}{\eta t_{\rm ev}}, \label{eq:dJsdt_full} \\
    \frac{\der e}{\der t} = &- \frac{1-e^2}{2e} \bigg[  1 - \frac{1}{k\sqrt{1-e^2}} \left( \tau_m \sin\psi + m \cos\psi \right)  \bigg] \frac{1}{\eta t_{\rm ev}}, \\
    \frac{\der \psi}{\der t} = &- \frac{1}{2 k \sqrt{1-e^2}} \bigg[ \tau_m \left( \frac{J_\ro}{J_\star} + \cos \psi \right) - m \sin \psi \bigg] \frac{1}{\eta t_{\rm ev}},
    \label{eq:obl_damp_full}
\end{align}
where
\begin{equation}
    \tau_m = \frac{\tau_m^+ + \tau_m^-}{2} = \frac{m \langle T_{x,m} \rangle}{\langle T_{z,m} \rangle} = \frac{k \Omega \langle T_{x,m} \rangle}{\langle \dot E_m \rangle}.
    \label{eq:tau_def}
\end{equation}
Depending on the eccentricity and obliquity, these equations predict the eccentricity and obliquity can damp or grow.


\section{Non-Linear Effects in the Radiative Cores of Low-Mass Stars}
\label{app:Nonlin}

As tidally-excited gravity waves propagate to the centers of low-mass stars, geometrical focusing causes their amplitudes to grow. This appendix considers how non-linearity may impact dissipation and backreact on resonant excitation. This is a complicated and uncertain subject, and we give here only a few rough calculations that are biased in favor of preserving  resonance locking. We consider both the non-linear theory of \cite{EssickWeinberg(2016)}, and our preferred alternative, tentative view.


\subsection{Non-Linear Breaking of the Three-Mode, Parametric Instability}
\label{app:Nonlin_inst}

When the amplitude of a `parent' gravity mode becomes sufficiently large,  two `child' standing-wave oscillations can be excited through a parametric instability \citep[e.g.][]{KumarGoldreich(1989), WuGoldreich(2001), Weinberg+(2012)}.  \cite{EssickWeinberg(2016)} found the child modes can  equilibrate more-or-less directly with the tidal forcing, leaving the parent mode's amplitude insensitive to distance from resonance. Accordingly, \cite{MaFuller(2021)} worried that resonance locking would be suppressed, particularly for relatively massive perturbers like hot Jupiters. In this subsection, we consider a simplified model of this parametric instability, reproducing the results of \cite{EssickWeinberg(2016)}, but also finding that the child standing modes can grow to such large amplitudes that they `break' (e.g.~\citealt{BarkerOgilvie(2010),BarkerOgilvie(2011)}).  The parent may thus be unable to rely on the child standing modes to balance the tidal forcing, in which case the resonant response of the parent may be preserved.

Our parent eigenmode displacement $\bxi_\ag$ oscillates with a frequency $\om_\ag$, and is non-linearly coupled to a set of child displacements $\bxi_\bg, \bxi_{\bg'}$, with oscillation frequencies $\om_\bg > \om_{\bg'}> 0$ 
and $\om_\bg + \om_{\bg'} \simeq \om_\ag$, through a coupling coefficient $\kappa_{\ag \bg \bg'}$ \citep[e.g.][]{WuGoldreich(2001), Schenk+(2001), Weinberg+(2012)}. 
We couple a child $\bxi_\bg$ to the parent $\bxi_\ag$ and a unique sibling $\bxi_{\bg'}$, 
and neglect additional couplings between child modes 
(see \citealt{Weinberg+(2012)} and \citealt{EssickWeinberg(2016)} for further discussion).  Picking a resonant ($\om_\ag \approx \om_{km}$) component of the tidal potential $\{k, \ell, m, m_\ro\}$, and normalizing the mode energy $\eps_\ag = E_\star = G M_\star^2/\Rs$, we write the parent-child equations of motion for the dimensionless mode amplitudes 
$c(t)$ as 
\begin{align}
    &\dot c_\ag + (\im \om_\ag + \cg_\ag) c_\ag = \im \om_\ag U^k_{\ell m m_\ro} I_{\ag \ell m} \e^{-\im \om_{km} t} + 2 \im \om_\ag \sum_{\bg} \kg_{\ag \bg \bg'} c_\bg c_{\bg'}, \nonumber \\
    &\dot c_\bg + (\im \om_\bg + \cg_\bg) c_\bg = 2 \im \om_\bg \kg_{\ag \bg \bg'} c_\ag c_{\bg'}^*,
    \hspace{8mm} 
    \dot c_{\bg'} + (\im \om_{\bg'} + \cg_{\bg'}) c_{\bg'} = 2 \im \om_\bg \kg_{\ag \bg \bg'} c_\ag c_{\bg}^* 
    \label{eq:dotca_nonlin}
\end{align}
for damping rates $\gamma$ (e.g. \citealt{Schenk+(2001), Weinberg+(2012)}; see also sections \ref{subsec:app_tide_pot} and \ref{subsec:app_mode_amp} for meanings of other variables). 
The full calculation of $\kappa_{\ag \bg \bg'}$ is challenging \citep[e.g.][]{VanBeeck+(2023)}.  For a $\sim$4 Gyr solar-mass star, 
using the WKB scaling for how parent and child modes behave close to the star's center, \cite{Weinberg+(2012), Weinberg+(2023)} show $\kappa_{\ag \bg \bg'}$ may be approximated as
\begin{equation}
    \kappa_{\ag \bg \bg'} \approx 2 \times 10^3 \left( \frac{P_\ag}{1 \ {\rm d}} \right)^2
    \label{eq:kappa_approx}
\end{equation}
where $P_\ag = 2\pi/\om_\ag$ is the oscillation period of the parent.  Tidal dissipation saps the parent's energy at the rate
\begin{equation}
    \dot E_{\rm tide} = -2 \om_\ag \Im \left(c_\ag^* U^k_{\ell m m_\ro} I_{\ag \ell m} \right)
    \label{eq:dotEtide}
\end{equation}
 where $\Im(A)$ denotes the imaginary part of $A$ \citep{Schenk+(2001), Weinberg+(2012), Burkart+(2014)}.

When the child modes are absent or only weakly excited 
($|c_\bg|, |c_{\bg'}| \ll |U^k_{\ell m m_\ro} I_{\ag \ell m}/(2\kappa_{\ag \bg \bg'})|^{1/2}$), $\dot E_{\rm tide}$ is dominated by `linear' dissipation from the parent:
\begin{equation}
E_\ag^{\rm lin} = |c_\ag|^2 E_\star = \frac{\om_\ag^2}{\Dg_\ag^2 + \cg_\ag^2} |U^k_{\ell m m_\ro} I_{\ag \ell m}|^2 E_\star,
\hspace{10mm}
\dot E_{\rm tide}^{\rm lin} = 2 \cg_\ag E_\ag^{\rm lin} = \frac{2 \cg_\ag \om_\ag^2 }{\Dg_\ag^2 + \cg_\ag^2} |U^k_{\ell m m_\ro} I_{\ag \ell m}|^2 E_\star 
\label{eq:dotE_lin}
\end{equation}
where $\Dg_\ag = \om_\ag - \om_{km}$ is the parent's de-tuning frequency. Equation (\ref{eq:dotE_lin}) describes the Lorentzian response of the parent which makes resonance locking possible. 

Parametric instability occurs when the parent's energy exceeds a certain threshold 
\begin{equation}
    E_\ag^{\rm thr, sw} = 
    \frac{1}{4 \kg_{\ag \bg \bg'}^2} \left( \frac{\cg_\bg^{\rm sw} \cg_{\bg'}^{\rm sw}}{\om_\bg \om_{\bg'}} \right) \left[ 1 + \left( \frac{\Delta_{\bg \bg'}}{\cg_\bg^{\rm sw} + \cg_{\bg'}^{\rm sw}} \right)^2 \right] E_\star 
    \label{eq:Ea_thr}
\end{equation}
where $\Delta_{\bg \bg'} = \om_\bg + \om_{\bg'} - \om_{km}$ is the child modes' de-tuning frequency, 
and $\cg_{\bg}^{\rm sw}, \cg_{\bg'}^{\rm sw}$ are the child standing-wave damping rates \citep{WuGoldreich(2001),Weinberg+(2012),EssickWeinberg(2016)}. Parametrically unstable child modes
can grow in amplitude and sap energy from the parent via non-linear interactions. If $E_\ag^{\rm lin} \gg E_\ag^{\rm thr, sw}$, the child mode energies $E_\bg$ and $E_\bg'$ reach 
\begin{equation}
    E_\bg^{\rm sw} = |c_\bg|^2 E_\star \simeq \sqrt{ \frac{\cg_{\bg'}^{\rm sw} \om_\bg}{\cg_\bg^{\rm sw} \om_{\bg'}} } \left| \frac{U^k_{\ell m m_\ro} I_{\ag \ell m}}{2 \kappa_{\ag \bg \bg'}} \right| E_\star, \hspace{10mm} E_{\bg'}^{\rm sw} = |c_{\bg'}|^2 E_\star \simeq \sqrt{ \frac{\cg_{\bg}^{\rm sw} \om_{\bg'}}{\cg_{\bg'}^{\rm sw} \om_{\bg}} } \left| \frac{U^k_{\ell m m_\ro} I_{\ag \ell m}}{2 \kappa_{\ag \bg \bg'}} \right| E_\star ,
    \label{eq:Eb_eq}
\end{equation}
respectively \citep{Weinberg+(2012)}, while the parent energy is limited to $E_\ag \approx E_\ag^{\rm thr, sw}$, which does not depend on $\Dg_\ag$. 

\begin{figure}
\centering
\includegraphics[width=\linewidth]{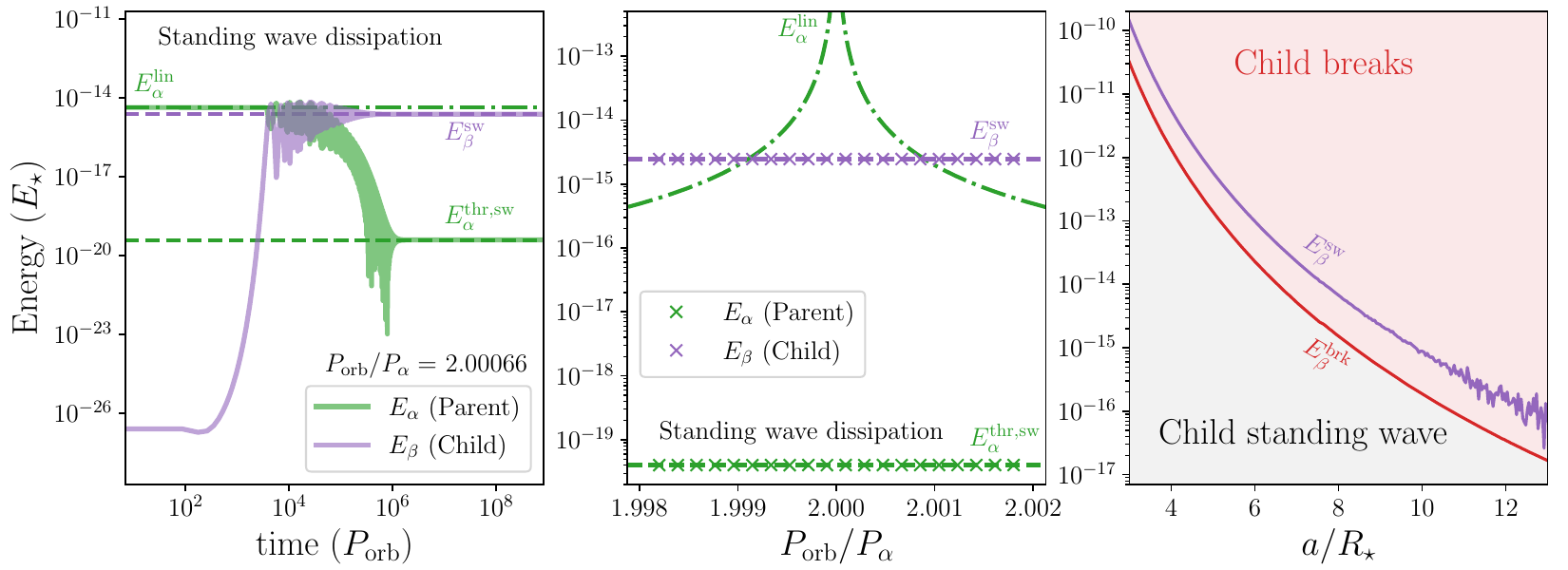}
\caption{
Parametric instability assuming the child modes behave as standing waves that do not break. We integrate equations ~\eqref{eq:dotca_nonlin} using the $U^{2}_{202}$ component of the gravitational potential of a Jupiter-mass planet on a circular ($e=0$), polar ($\psi = 90^\circ$) orbit of semi-major axis $a$. The parent stellar g mode, of oscillation period $P_\ag = 1.5 \ \der$, is computed from \texttt{GYRE} using a 4-Gyr-old, solar-mass star of radius $R_\star$ modeled with \texttt{MESA}. Equations \eqref{eq:dotca_nonlin} are integrated to a time $t_{\rm end} = 1.1 \times 10^{10} \ \om_\ag^{-1}$, with $c_\ag(0)$ set by eq.~\eqref{eq:ca_force}, and $c_\bg(0) = 10^{-6} c_\ag(0)$.
\textit{Left}: Energies of the parent mode and one of the twin child modes vs.~time, for frequency de-tuning $\Dg_\ag = 3.3 \times 10^{-4} \om_\ag$. Parent and child energies evolve to their equilibrium standing-wave energies $E_\ag^{\rm thr,sw}$ (eq.~\ref{eq:Ea_thr}) and $E_\bg^{\rm sw}$ (eq.~\ref{eq:Eb_eq}), respectively. 
\textit{Middle}: Parent and child energies (crosses) at times $t > t_{\rm end}/2$, vs.~forcing frequency.  Both parent and child energies are independent of $\Dg_\ag$, in contrast to the Lorentzian resonant response of the parent mode neglecting child modes ($E_\ag^{\rm lin}$, eq.~\ref{eq:dotE_lin}).  
\textit{Right}: The child mode energy $E_\bg^{\rm sw}$ exceeds the child breaking energy $E_\bg^{\rm brk}$ (eq.~\ref{eq:Eb_max}) for values of $a/R_\star$ relevant to hot Jupiter systems. Breaking of standing child modes might help restore the resonant response of the parent mode---see Figs.~\ref{fig:E_mode_brk} and \ref{fig:tmode}.
\label{fig:E_mode_nl}
}
\end{figure}

To demonstrate how parametric instability and the spawning of child modes can suppress the resonant response of the parent, we solve equations (\ref{eq:dotca_nonlin}) using the following model. We consider tidal forcing from the $\{k,l,m, m_{\rm orb}\} = \{2,2,0,2\}$ component of $U$ of a Jupiter-mass planet on a circular and polar orbit with period $P_{\rm orb}$. The parent mode oscillation period is taken to be $P_\alpha = 1.5$ days, and the child oscillations are assumed to be `twins,' with nearly equal frequencies in perfect parametric resonance with the parent 
($\om_\bg = \om_{\bg'} = \om_{km}/2$, $c_\bg = c_{\bg'}$, $\Dg_{\bg \bg'} = 0$). 
Because twin standing waves minimize $E_\ag^{\rm thr, sw}$, they are the child modes most likely to be parametrically excited \citep{WuGoldreich(2001),EssickWeinberg(2016)}. The stellar oscillation code \texttt{GYRE} is used to calculate the parent modes of a 4-Gyr-old, solar-mass star modeled with \texttt{MESA}, with parent radiative damping rates $\cg_\ag$ evaluated using the \texttt{GYRE} inlists in \cite{Fuller(2017)}.  The damping rates of standing child modes are approximately $\cg_\bg^{\rm sw}, \cg_{\bg'}^{\rm sw} \approx 4 \cg_\ag$ (for angular degrees $\ell_\bg, \ell_{\bg'} = 2$; \citealt{Weinberg+(2012)}). Numerical solutions to equations (\ref{eq:dotca_nonlin}) are shown in Figure~\ref{fig:E_mode_nl},
using \texttt{odeint} from \texttt{scipy} with $\texttt{rtol}=10^{-12}$, $\texttt{atol}=10^{-16}$ \citep[also getting rid of fast oscillations through a variable transform, see e.g.][for details]{EssickWeinberg(2016)},
whose 
left and middle panels illustrate how child mode energies eventually dominate parent mode energies, and how both energies are independent of the frequency detuning. Resonance locking would appear to be impossible when standing-wave child modes are fully excited.

\begin{figure}
\centering
\includegraphics[width=\linewidth]{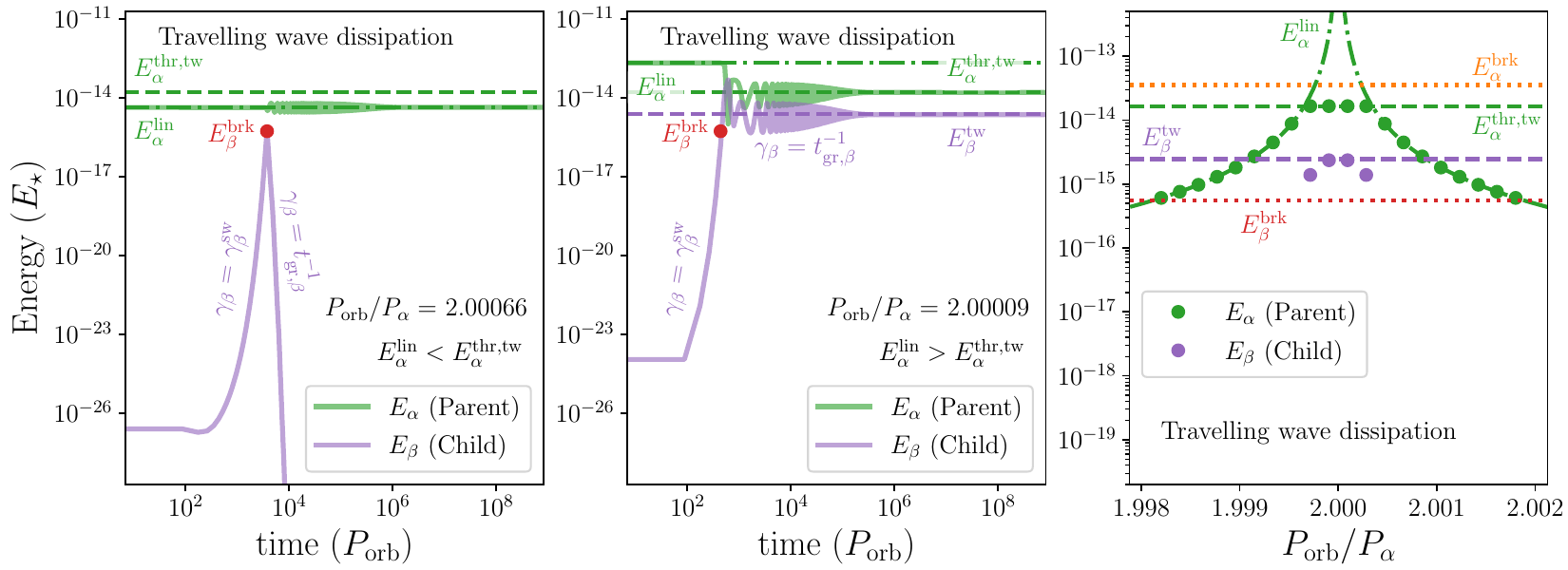}
\caption{
Same as Fig.~\ref{fig:E_mode_nl}, except we replace $\cg_\bg$ with $t_{{\rm gr},\bg}^{-1}$ (eq.~\ref{eq:ta_gr}) once $E_\bg > E_\bg^{\rm brk}$ and child modes become traveling waves.
\textit{Left}: Energies of the parent (solid green line) and child (solid purple line) vs.~time for $\Dg_\ag = 3.3 \times 10^{-4} \om_\ag$. 
For these parameters, the parent's energy stays below the threshold  $E_\ag^{\rm thr,tw}$ (eq.~\ref{eq:Ea_tw}) for maintaining a traveling child wave. Thus the child mode disappears as soon as it breaks, and the parent energy remains close to $E_\ag^{\rm lin}$. \textit{Middle}: Same as left but for $\Dg_\ag = 4.7 \times 10^{-5} \om_\ag$ (closer to exact resonance). Here the parent energy exceeds $E_\ag^{\rm thr,tw}$ and can sustain a traveling child mode. The parent and child energies asymptote to $E_\ag^{\rm thr,tw}$ and $E_\bg^{\rm tw}$ (eq.~\ref{eq:Eb_tw}), respectively. 
\textit{Right}: Equilibrium parent and child energies (evaluated at $t > t_{\rm end}/2$) vs.~forcing frequency. Exact resonance is located at $P_{\rm orb}/P_\alpha = 2$. Away from exact resonance, child energies are too low to be plotted, and leave parent energies near $E_\ag^{\rm lin}$. Closer to exact resonance, child energies grow to $E_\bg^{\rm tw}$, causing the parent's energy to flat-line at $E_\ag^{\rm thr,tw}$, and preventing the parent from breaking.  
\label{fig:E_mode_brk}
}
\end{figure}

What is neglected in the above discussion is the possibility that a mode amplitude can become large enough to render the background medium, which is ordinarily stably stratified, convectively unstable. When this occurs, the mode is said to `overturn' or `break', and the oscillation reduces to a single travelling wave, rather than a standing wave (= oppositely directed travelling waves trapped in a resonant cavity).
Denoting $\xi_{r} = \iint  \bxi \bcdot {\bm {\hat r}} Y_{\ell m}^* \sin \theta \, \der \theta \, \der \vphi$, we have that a mode with amplitude $c > c^{\rm brk}$ may break, where 
\begin{equation}
    c^{\rm brk} \left| \frac{\der \xi_{r}}{\der r} \right|_{\max} \approx 1
    \label{eq:dxirdr_le_1}
\end{equation}
and $|\der \xi_{r}/\der r|_{\max}$ is the maximum amplitude of $\der \xi_{r}/\der r$ over the star's radial extent \citep[e.g.][]{GoodmanDickson(1998),BarkerOgilvie(2010),BarkerOgilvie(2011),Barker(2011),Guo+(2023)}.  
Because the mode breaking energy $E^{\rm brk} = |c^{\rm brk}|^2 E_\star \propto \om^6$ \citep{EssickWeinberg(2016)}, the child mode breaking energies $E_\bg^{\rm brk}$ and $E_{\bg'}^{\rm brk}$ can be scaled to the parent's:
\begin{equation}
    E_\bg^{\rm brk} \approx \left( \frac{\om_\bg}{\om_\ag} \right)^6 E_\ag^{\rm brk}, 
    \hspace{10mm}
    E_{\bg'}^{\rm brk} \approx \left( \frac{\om_{\bg'}}{\om_\ag} \right)^6 E_\ag^{\rm brk}.
    \label{eq:Eb_max}
\end{equation}
Because $\om_\bg, \om_{\bg'} < \om_\ag$, the child breaking energy is always smaller than that of the parent.
In the right panel of Figure \ref{fig:E_mode_nl}, we see that for our standing-wave twin-child model, $E_\bg^{\rm sw} > E_\bg^{\rm brk}$ and $E_{\bg'}^{\rm sw} > E_{\bg'}^{\rm brk}$ (albeit only by factors of several), for the entire range of $a/\Rs$ tested. The breaking of standing child waves opens the possibility of restoring the resonant response of the parent.

Once broken, child modes become traveling waves, and may no longer be able to sap the parent of energy as effectively as standing waves do. Travelling waves rapidly deplete their energy over their group crossing time 
\begin{align}
    t_{{\rm gr}} \simeq \frac{\pi n}{\om} 
\label{eq:ta_gr}
\end{align}
\citep[e.g.][]{ZanazziWu(2021), MaFuller(2021)}, where $n$ denotes the number of radial nodes.  The parent-child coupling coefficient $\kappa_{\ag \bg \bg'}$ is largely the same whether the child mode is standing or travelling \citep{Weinberg+(2012)}.  However, to parametrically excite a travelling child, the parent's energy must now exceed a much higher threshold 
\begin{equation}
    E_\ag^{\rm thr,tw}  = \frac{1}{4 \kappa_{\ag \bg \bg'}^2}
 \left( \frac{\cg_\bg^{\rm tw} \cg_{\bg'}^{\rm tw}}{\om_\bg \om_{\bg'}} \right) E_\star 
 \label{eq:Ea_tw}
\end{equation}
\citep{WuGoldreich(2001)}, where the child travelling wave damping rate is $\cg_\bg^{\rm tw}, \cg_{\bg'}^{\rm tw} \approx t_{\rm gr,\bg}^{-1}, t_{\rm gr, \bg'}^{-1} \approx (4 t_{\rm gr, \ag})^{-1}$ (for $\ell_\bg, \ell_{\bg'} = 2$). The number of parent radial nodes can be calculated from \texttt{GYRE}.

Figure \ref{fig:E_mode_brk} presents new solutions to equations (\ref{eq:dotca_nonlin}), now taking $\gamma_\beta = \cg_\bg^{\rm tw}$ when $E_\bg > E_\bg^{\rm brk}$.  For $P_{\rm orb}/P_\alpha = 2.00066$, we have $E_\ag^{\rm thr,tw} > E_\ag >E_\ag^{\rm thr,sw}$: standing child modes are parametrically excited but die upon breaking (Fig.~\ref{fig:E_mode_brk}, left panel). Closer to exact resonance ($P_{\rm orb}/P_\alpha = 2.00009$), we have $E_\ag >E_\ag^{\rm thr,tw}$: now each child traveling mode grows to an energy
\begin{equation}
    E_\bg^{\rm tw} = |c_\bg|^2 E_\star \approx \left| \frac{U^k_{\ell m m_\ro} I_{\ag \ell m}}{2 \kappa_{\ag \bg \bg'}} \right| E_\star ,
\label{eq:Eb_tw}
\end{equation}
while the parent depletes in energy to $E_\ag^{\rm thr, tw}$ (Fig.~\ref{fig:E_mode_brk}, middle panel).  The resonant response of the parent mode is restored to an extent (Fig.~\ref{fig:E_mode_brk}, right panel); moreover, while the child mode breaks, the parent mode does not (although the margin of safety is only a factor of a few between $E_\ag^{\rm thr, tw}$ and $E_\ag^{\rm brk}$). 
Comparing $E_\ag^{\rm thr,tw}$ to $E_\ag^{\rm brk}$ over a broad range of $a/R_\star$, parent modes might excite travelling children, without the parents breaking themselves (Fig.~\ref{fig:tmode}, right panel; with a factor of a few between $E_\ag^{\rm thr,tw}$ and $E_\ag^{\rm brk}$).
In the next subsection \ref{app:Nonlin_diss}, we will continue to study this `traveling child' scenario and its implications for resonance locking.

\cite{EssickWeinberg(2016)} argued that child mode breaking can be avoided by accounting for `grandchild' modes, since child modes are themselves parametrically unstable.  We question this argument.  
From their Fig.~7, 
the typical energies of grandchild modes ($E_{\rm grandchild} \sim 10^{-11} E_\star$ for $a/\Rs \approx 4$, and $E_{\rm grandchild} \sim 10^{-15} E_\star$ for $a/\Rs \approx 9$) appear $\sim$$10^{2}$ times larger than their respective breaking energies 
(estimated using $E_\beta^{\rm brk}$ from the right panel of our Fig.~\ref{fig:E_mode_nl}, and the scaling $E_{\rm grandchild}^{\rm brk} \sim E_{\beta}^{\rm brk} / 2^6$ from eq.~\ref{eq:Eb_max}). Thus grandchild modes would seem even more prone to breaking than child modes. 

\subsection{Resonant Dissipation from Travelling Wave Child Modes}
\label{app:Nonlin_diss}

\begin{figure}
\centering
\includegraphics[width=\linewidth]{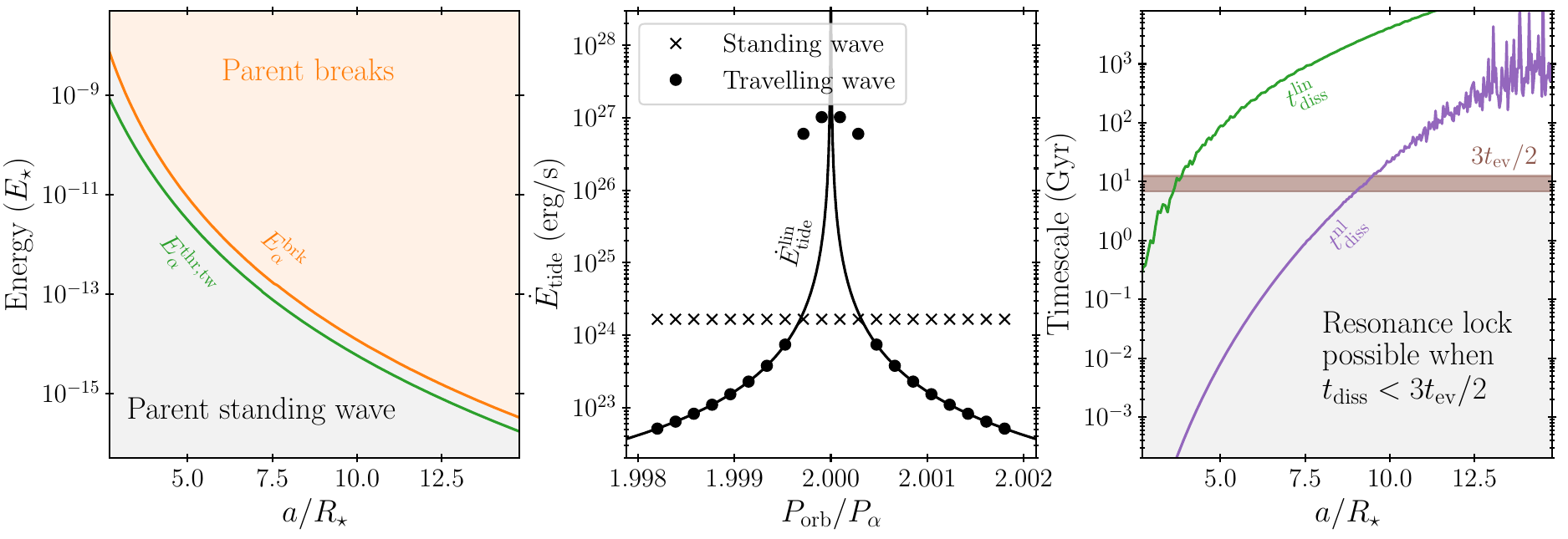}
\caption{
Parent mode amplitudes, energy dissipation rates, and dissipation timescales, using the same hot Jupiter + star model of Fig.~\ref{fig:E_mode_brk}.
\textit{Left}:  Comparing the threshold energy $E_\ag^{\rm thr, tw}$
(eq.~\ref{eq:Ea_tw}) needed for a parent to excite a traveling child
mode, to the parent breaking energy $E_\ag^{\rm brk}$. The parent may
parametrically excite travelling wave child modes without itself
breaking. 
\textit{Middle}: Average tidal dissipation rate
(eq.~\ref{eq:dotEtide}) at times $t > t_{\rm end}/2$, either assuming the
child modes behave as standing waves (crosses; see also
Fig.~\ref{fig:E_mode_nl}), or accounting for how child standing waves
break to become traveling waves (solid circles; see also
Fig.~\ref{fig:E_mode_brk}). There is no resonant response when child
modes are standing waves, but there is when child modes become
traveling waves. \textit{Right}: Non-linear dissipation timescale
$t^{\rm nl}_{\rm diss}$ (purple) compared to the stellar evolutionary
timescale $3 t_{\rm ev}/2$ (brown bar, derived from the $M_\star = 1\
{\rm M}_{\odot}$ curve from Fig.~\ref{fig:aveN_t}, top panel).  Non-linear damping may sustain resonance locks out to $(a/R_\star)_{\rm crit} \sim 8-10$.  For comparison, the timescale $t^{\rm lin}_{\rm diss}$ assuming linear damping is also shown. 
}
\label{fig:tmode}
\end{figure}

  For a hot Jupiter to resonantly lock onto a stellar mode, the tidal
  dissipation time $t_{\rm diss} = |E_{\rm orb}|/|\dot E_{\rm tide}|$
  must be shorter than the stellar evolution time $3 t_{\rm ev}/2$.
  Following appendix~\ref{app:Nonlin_inst}, when traveling wave child modes are
  excited, they dominate the dissipation rate near exact resonance at
  $P_{\rm orb} = 2 P_\ag$ (Fig.~\ref{fig:tmode}, middle panel,
  solid circles near figure center).  Though the parent energy exceeds that of
  the child modes (Fig.~\ref{fig:E_mode_brk}, right panel), the
  timescale $t_{\rm gr,\bg}$ over which child traveling waves dissipate is much shorter than
  the dissipation time $1/\gamma_\ag$ for the parent; consequently, the energy
  dissipation rate from the two child modes exceeds that from the parent
  (i.e., $2 \times 2 t_{\rm gr,\bg}^{-1} E_\bg^{\rm tw} > 2 \cg_\ag E_\ag^{\rm thr,tw}$).
  Thus the tidal dissipation time (superscript ``nl'' for non-linear,
  following the main text)  is given by
\begin{equation}
    t_{\rm diss}^{\rm nl} = \frac{|E_{\rm orb}|}{4 t_{\rm gr,\bg}^{-1} E_\bg^{\rm tw}}
 .   \label{eq:tmin_nl}
\end{equation}
Comparing $t_{\rm diss}^{\rm nl}$ (evaluated for a hot Jupiter on a polar orbit $\psi = 90^\circ$) to $3 t_{\rm ev}/2$ in the right panel of Fig.~\ref{fig:tmode}, we see that
resonance locks are possible for semi-major axes inside 
$(a/R_\star)_{\rm crit} \sim 8-10$.

Travelling wave child energies $E_\bg^{\rm tw}$ are proportional to the tidal
potential $|U^k_{\ell m m_\ro}|$. In our
model we focus on $|U^2_{202}| $, which scales as $\sin^2\psi$.
Thus $t_{\rm diss}^{\rm nl} \propto 1/E_\bg^{\rm tw} \propto
1/\sin^2\psi$. At a given $a/R_\star < (a/R_\star)_{\rm crit}$,
resonance locks are possible for $\sin \psi = 1$ through 
\begin{equation}
    |\sin \psi_{\rm unlock}^{\rm nl}| \approx \left( \frac{2 t_{{\rm
            diss},\psi=90^\circ}^{\rm nl}}{3 t_{\rm ev}} \right)^{1/2}
    . 
    \label{eq:psi_min_nl}
\end{equation}
The purple curves in Figure~\ref{fig:psi_min} plot
equation~\eqref{eq:psi_min_nl} for $t_{\rm ev} = \langle N \rangle / \langle \dot N \rangle = 4.5-8.3$ Gyr (Fig.~\ref{fig:aveN_t}, $M_\star = 1 \ {\rm M}_\odot$),
using $t^{\rm nl}_{\rm diss, \psi=90^\circ}$ from
Fig.~\ref{fig:tmode}. The region to the left of the purple curves in
Fig.~\ref{fig:psi_min} allows for resonance
locking and is seen to contain a number of hot-Jupiter-hosting
cool stars, many having low obliquities.


For reference, ``linear'' (radiative) damping of a standing-wave parent
mode can dissipate energy over a timescale
\begin{equation}
    t_{\rm diss}^{\rm lin} = \frac{|E_{\rm orb}|}{2 \cg_\ag E_\ag^{\rm brk}}
 .   \label{eq:tmin_lin}
\end{equation}
Equation (\ref{eq:tmin_lin}) is a minimum timescale because it uses the maximum
parent mode energy $E_\ag^{\rm brk}$, above which the parent mode
breaks (and which our model approaches to within a factor of a few;
see Fig.~\ref{fig:E_mode_brk}, right panel). Linear dissipation can sustain resonance locks out to 
$(a/R_\star)_{\rm crit} \sim 4$ (Fig.~\ref{fig:tmode}, right
panel). Since $E_\ag^{\rm brk}$ depends only on the core's
stratification, $\min t_{\rm diss}^{\rm lin}$ is independent of
obliquity, predicting resonance locks may hold until $\psi = 0$ and
the tidal torque vanishes (Fig.~\ref{fig:ModeEx}, lower left panel). 
In Fig.~\ref{fig:psi_min} we plot $\psi_{\rm unlock}^{\rm lin}$ as a vertical green bar at $(a/R_\star)_{\rm crit} \sim 4$; linear dissipation appears insufficient to align most low-mass stars hosting hot Jupiters.

\bibliography{main_v3}
\bibliographystyle{aasjournal}



\end{document}